\documentclass[twocolumn,times]{aastex63}
\usepackage{natbib,multirow}
\usepackage[caption=false]{subfig}

\newcommand{\kms}{km\,s$^{-1}$}
\newcommand{\cii}{[\ion{C}{2}]}

\newcommand{\lfir}{$L_{\mathrm{FIR}}$}
\newcommand{\ltir}{$L_{\mathrm{TIR}}$}
\newcommand{\lcii}{$L_\mathrm{[CII]}$}

\newcommand{\lsun}{$L_\sun$}
\newcommand{\msun}{$M_\sun$}

\newcommand{\msunyr}{$M_\sun$\,yr$^{-1}$}

\received{July 17, 2020}
\revised{October 5, 2020}
\accepted{October 23, 2020}

\shorttitle{ALMA kpc \cii\ and dust continuum imaging of 27 $z\sim6$ quasars}
\shortauthors{Venemans et al.}

\begin{document}

\title{Kiloparsec-scale ALMA Imaging of \cii\ and Dust Continuum Emission of 27 Quasar Host Galaxies at $z\sim6$}

\correspondingauthor{Bram P.\ Venemans}
\email{venemans@mpia.de}

\author[0000-0001-9024-8322]{Bram P.\ Venemans}
\affiliation{Max-Planck Institute for Astronomy, K{\"o}nigstuhl 17, D-69117 Heidelberg, Germany}

\author[0000-0003-4793-7880]{Fabian Walter}
\affiliation{Max-Planck Institute for Astronomy, K{\"o}nigstuhl 17, D-69117 Heidelberg, Germany}
\affiliation{National Radio Astronomy Observatory, Pete V. Domenici Array Science Center, P.O. Box 0, Socorro, NM 87801, USA}

\author[0000-0002-9838-8191]{Marcel Neeleman}
\affiliation{Max-Planck Institute for Astronomy, K{\"o}nigstuhl 17, D-69117 Heidelberg, Germany}

\author[0000-0001-8695-825X]{Mladen Novak}
\affiliation{Max-Planck Institute for Astronomy, K{\"o}nigstuhl 17, D-69117 Heidelberg, Germany}

\author[0000-0003-3191-9039]{Justin Otter}
\affiliation{Max-Planck Institute for Astronomy, K{\"o}nigstuhl 17, D-69117 Heidelberg, Germany}
\affiliation{Department of Physics \& Astronomy, Johns Hopkins University, Bloomberg Center, 3400 N. Charles St., Baltimore, MD 21218, USA}

\author[0000-0002-2662-8803]{Roberto Decarli}
\affiliation{INAF -- Osservatorio di Astrofisica e Scienza dello Spazio di Bologna, via Gobetti 93/3, I-40129, Bologna, Italy}

\author[0000-0002-2931-7824]{Eduardo Ba\~nados}
\affiliation{Max-Planck Institute for Astronomy, K{\"o}nigstuhl 17, D-69117 Heidelberg, Germany}

\author[0000-0002-0174-3362]{Alyssa Drake}
\affiliation{Max-Planck Institute for Astronomy, K{\"o}nigstuhl 17, D-69117 Heidelberg, Germany}

\author[0000-0002-6822-2254]{Emanuele Paolo Farina}
\affiliation{Max Planck Institut f\"ur Astrophysik, Karl--Schwarzschild--Stra{\ss}e 1, D-85748, Garching bei M\"unchen, Germany}

\author[0000-0002-1173-2579]{Melanie Kaasinen}
\affiliation{Max-Planck Institute for Astronomy, K{\"o}nigstuhl 17, D-69117 Heidelberg, Germany}

\author[0000-0002-5941-5214]{Chiara Mazzucchelli}
\affiliation{European Southern Observatory, Alonso de C\'{o}rdova 3107, Vitacura, Regi\'{o}n Metropolitana, Chile}

\author[0000-0001-6647-3861]{Chris Carilli}
\affiliation{National Radio Astronomy Observatory, Pete V. Domenici Array Science Center, P.O. Box 0, Socorro, NM 87801, USA}

\author[0000-0003-3310-0131]{Xiaohui Fan}
\affiliation{Steward Observatory, University of Arizona, 933 N Cherry Ave, Tucson, AZ 85719, USA}

\author[0000-0003-4996-9069]{Hans-Walter Rix}
\affiliation{Max-Planck Institute for Astronomy, K{\"o}nigstuhl 17, D-69117 Heidelberg, Germany}

\author[0000-0003-4956-5742]{Ran Wang}
\affiliation{Kavli Institute for Astronomy and Astrophysics, Peking University, No.5 Yiheyuan Road, Haidian District, Beijing, 100871, China}

\begin{abstract}
We present a study of the \cii\ 158\,$\mu$m line and underlying far-infrared (FIR) continuum emission of 27 quasar host galaxies at $z\sim6$, traced by the Atacama Large Millimeter/submiilimeter Array at a spatial resolution of $\sim$1\,physical kpc. The \cii\ emission in the bright, central regions of the quasars have sizes of 1.0--4.8\,kpc. The dust continuum emission is typically more compact than \cii. We find that 13/27 quasars (approximately one-half) have companion galaxies in the field, at projected separations of 3--90\,kpc. The position of dust emission and the Gaia-corrected positions of the central accreting black holes are cospatial (typical offsets $\lesssim$0\farcs1). This suggests that the central black holes  are located at the bottom of the gravitational wells of the dark matter halos in which the $z>6$ quasar hosts reside. Some outliers with offsets $\sim$500\,pc can be linked to disturbed morphologies, most likely due to ongoing or recent mergers. We find no correlation between the central brightness of the FIR emission and the bolometric luminosity of the accreting black hole. The FIR-derived star-formation rate densities (SFRDs) in the host galaxies peak at the galaxies' centers, at typical values between 100 and 1000\,\msunyr\,kpc$^{-2}$. These values are below the Eddington limit for star formation, but similar to those found in local ultraluminous infrared gaalxies. The SFRDs drop toward larger radii by an order of magnitude. Likewise, the \cii/FIR luminosity ratios of the quasar hosts are lowest in their centers (few $\times 10^{-4}$) and increase by a factor of a few toward the galaxies' outskirts, consistent with resolved studies of lower-redshift sources.
\end{abstract}


\section{Introduction} 
\label{sec:introduction}

Luminous quasars at high redshift, $z\gtrsim6$, are extreme astrophysical objects. Accretion onto their central supermassive black holes with masses exceeding $10^9$\,\msun\ \citep[e.g.,][]{der14,maz17b,she19} results in extremely bright emission in the rest-frame ultraviolet bands \citep[absolute magnitudes up to $M_\mathrm{UV} > -29$; see, e.g.,][]{wu15}. This means that even the most distant quasars currently known at $z=7.5$ \citep{ban18a,yan20} can be detected with a 4\,m class near-infrared telescope in a few minutes. 

Similarly, it has been demonstrated that the galaxies that harbor the central supermassive black holes can also be very luminous in the rest-frame far-infrared (FIR). The \cii\ emission line, one of the main coolants of the interstellar medium (ISM), and the underlying dust continuum emission can be detected with state-of-the-art millimeter facilities, such as the Atacama Large Millimeter/submillimeter Array (ALMA) and the NOrthern Extended Millimeter Array (NOEMA), in less than an hour \citep[e.g.,][]{dec18,yan19}. The global \cii\ and dust properties of the host galaxies of a large sample of luminous $z\sim6$ quasars with central black holes with masses $\gtrsim10^9$\,\msun\ were recently presented in \citet{dec18} and \citet{ven18}. In general, the \cii\ emission in these quasar host galaxies has a luminosity in the range $10^9-10^{10}$\,\lsun, from which \citet{dec18} infer star-formation rates (SFRs) of $\sim$200--2000\,\msunyr.  Based on the detection of the dust emission, \citet{ven18} derived FIR luminosities of \lfir\,$>10^{12}$\,\lsun\ for the majority of quasar hosts and high dust masses of $10^7-10^9$\,\msun. The host galaxies of distant quasars thus provide a unique opportunity to study the formation and build-up of massive galaxies in the early universe in detail. 

One of the surprising results of the FIR imaging of distant quasar host galaxies was the discovery of massive, gas-rich companion galaxies within 60\,kpc and 600\,\kms\ of the quasar \citep{wan11a,dec17,wil17,nee19,ven19}. Similar gas-rich galaxies have been discovered around luminous quasars at slightly lower redshift, $z\sim5$ \citep[e.g.,][]{tra17,ngu20}. The fraction of quasars that show at least one nearby galaxy at a similar redshift is at least 15\%--30\% \citep{dec17,ngu20}, indicating that major mergers might play a significant role in triggering these luminous quasars, in contrast to (some) studies at lower redshifts \citep[$z\sim2$, e.g.,][]{mar19}. However, signatures of ongoing or recent merging events in high-redshift quasar host galaxies have been rarely observed \citep[e.g.,][]{ban19,dec19,ven19}, mostly due to the limited spatial resolution of the FIR observations.

So far, millimeter facilities have observed a small sample of quasar host galaxies at high-enough spatial resolution to resolve the emission of the quasar host galaxies. Despite the small number of targets, there appears to be a surprising variety in the morphology of gas and dust emission \citep[e.g.,][]{wal09b,sha17,nee19,ven17a,ven19}. Some objects show very compact ($<$1.5\,kpc) dispersion-dominated emission, with implied high central gas densities and star formation surface densities \citep[e.g.,][]{ven17a,nee19}. Others show more extended disk-like, rotation-dominated \cii\ emission, distributed over many kpc$^2$ \citep[e.g.,][]{wan13,sha17,pen20}. In other cases, the \cii\ emission does not show ordered rotation, but rather a very disturbed morphology, possibly due to interaction with nearby galaxies \citep[e.g.,][]{dec19,ven19}. These vastly different morphologies and gas kinematics could provide insights into early black hole growth (i.e., is the accretion onto the central black hole related to whether or not the gas kinematics are disturbed in the central region of the host galaxy?).

We therefore initiated a comprehensive study of $z\sim6$ quasar host galaxies using all available high-spatial-resolution FIR observations to probe the gas and dust on scales of $\sim$1\,kpc. The main goals of this study are to (1) determine the morphology, size, and kinematics of the \cii-emitting region and of the continuum emission in order to differentiate between different merger states and to constrain the dynamical mass of quasar hosts; (2) resolve the FIR surface density on kiloparsec scales to determine whether the central active galactic nucleus (AGN) plays a role in the heating of the dust; and (3) accurately measure the profile and extent of the \cii\ emission to explore the presence of possible gas infall or outflows.

These goals will be addressed in three parallel papers. In this paper, the first of the series, we will outline the sample and  present new and archival high-spatial-resolution ALMA observations of $z\sim6$ quasar host galaxies. We will discuss the global properties of the sample, study the size and morphology of the gas and dust emission, and examine the environment. In another paper of the series, \citet{nov20} investigate the presence of large-scale spatial features in the \cii\ and dust emission and constrain the broad spectral features in the \cii\ emission line to look for evidence of gas outflows. In a third paper, Neeleman M.\ et al.\ (2020, in preparation) analyze the kinematics of the \cii\ emission in the quasar hosts and constrain the dynamical mass of the galaxies. They further investigate the relation between the dynamical mass of the quasar host galaxies and the mass of the central black hole.

\begin{deluxetable*}{lcccccccc}
\tablecaption{Characteristics of the $z>5.7$ Quasar Sample and the ALMA Observations. \label{tab:sample}}
\tablewidth{0pt}
\tablehead{\colhead{Name} & 
\colhead{R.A.\ (ICRS)\tablenotemark{a}} &
\colhead{Decl.\ (ICRS)\tablenotemark{a}} &
\colhead{$\Delta$R.A.\,$\times$\,$\Delta$Decl.\tablenotemark{a}} &
\colhead{Project ID} & 
\colhead{$t_\mathrm{on-source}$\tablenotemark{b}} & 
\colhead{$n_\mathrm{ant}$\tablenotemark{c}} &
\colhead{Beam\tablenotemark{d}} &
\colhead{rms per 30\,MHz} \\
\colhead{} & 
\colhead{} & 
\colhead{} &
\colhead{} & 
\colhead{} & 
\colhead{(min.)} & 
\colhead{} &
\colhead{} &
\colhead{(mJy\,beam$^{-1}$)}}
\startdata
P007+04 & 00$^\mathrm{h}$28$^\mathrm{m}$06$.\!\!^\mathrm{s}$568 & +04$^\circ$57$^\prime$25$\farcs$39 & 0\farcs04\,$\times$\,0\farcs05 & 2017.1.01301.S & 29 & 44 & 0\farcs26$\times$0\farcs22 & 0.25 \\
P009--10 & 00$^\mathrm{h}$38$^\mathrm{m}$56$.\!\!^\mathrm{s}$519 & --10$^\circ$25$^\prime$53$\farcs$97 & 0\farcs04\,$\times$\,0\farcs04 & 2017.1.01301.S & 26 & 44 & 0\farcs32$\times$0\farcs22 & 0.54 \\
J0100+2802 & 01$^\mathrm{h}$00$^\mathrm{m}$13$.\!\!^\mathrm{s}$025 & +28$^\circ$02$^\prime$25$\farcs$80 & 0\farcs03\,$\times$\,0\farcs02 & 2015.1.00692.S & 72 & 44 & 0\farcs24$\times$0\farcs12 & 0.18 \\
\multirow{2}{*}{J0109--3047} & \multirow{2}{*}{01$^\mathrm{h}$09$^\mathrm{m}$53$.\!\!^\mathrm{s}$136} & \multirow{2}{*}{--30$^\circ$47$^\prime$26$\farcs$30} & \multirow{2}{*}{0\farcs02\,$\times$\,0\farcs05} & 2013.1.00273.S & 36 & 40 & \multirow{2}{*}{0\farcs20$\times$0\farcs17} & \multirow{2}{*}{0.20} \\
{} & {} & {} & {} & 2015.1.00399.S & 32 & 41 & {} & {} \\
J0129--0035 & 01$^\mathrm{h}$29$^\mathrm{m}$58$.\!\!^\mathrm{s}$515 & --00$^\circ$35$^\prime$39$\farcs$81 & 0\farcs03\,$\times$\,0\farcs03 & 2012.1.00240.S & 76 & 37--41 & 0\farcs22$\times$0\farcs16 & 0.22 \\
J025--33 & 01$^\mathrm{h}$42$^\mathrm{m}$43$.\!\!^\mathrm{s}$720 & --33$^\circ$27$^\prime$45$\farcs$61 & 0\farcs06\,$\times$\,0\farcs10 & 2017.1.01301.S & 24 & 44 & 0\farcs24$\times$0\farcs21 & 0.27 \\
P036+03 & 02$^\mathrm{h}$26$^\mathrm{m}$01$.\!\!^\mathrm{s}$873 & +03$^\circ$02$^\prime$59$\farcs$24 & 0\farcs04\,$\times$\,0\farcs03 & 2015.1.00399.S & 75 & 41--42 & 0\farcs15$\times$0\farcs12 & 0.18 \\
\multirow{2}{*}{J0305--3150} & \multirow{2}{*}{03$^\mathrm{h}$05$^\mathrm{m}$16$.\!\!^\mathrm{s}$918} & \multirow{2}{*}{--31$^\circ$50$^\prime$55$\farcs$85} & \multirow{2}{*}{0\farcs01\,$\times$\,0\farcs04} & 2013.1.00273.S & 16 & 34 & \multirow{2}{*}{0\farcs18$\times$0\farcs15} & \multirow{2}{*}{0.26} \\
{} & {} & {} & {} & 2015.1.00399.S & 38 & 44 & {} & {} \\
P065--26 & 04$^\mathrm{h}$21$^\mathrm{m}$38$.\!\!^\mathrm{s}$050 & --26$^\circ$57$^\prime$15$\farcs$72 & 0\farcs05\,$\times$\,0\farcs06 & 2017.1.01301.S & 25 & 47 & 0\farcs29$\times$0\farcs21 & 0.32 \\
J0842+1218 & 08$^\mathrm{h}$42$^\mathrm{m}$29$.\!\!^\mathrm{s}$438 & +12$^\circ$18$^\prime$50$\farcs$47 & 0\farcs03\,$\times$\,0\farcs02 & 2016.1.00544.S & 54 & 42 & 0\farcs27$\times$0\farcs23 & 0.28 \\
J1044--0125 & 10$^\mathrm{h}$44$^\mathrm{m}$33$.\!\!^\mathrm{s}$040 & --01$^\circ$25$^\prime$02$\farcs$08 & 0\farcs01\,$\times$\,0\farcs01 & 2012.1.00240.S & 76 & 34--35 & 0\farcs22$\times$0\farcs16 & 0.39 \\
J1048--0109 & 10$^\mathrm{h}$48$^\mathrm{m}$19$.\!\!^\mathrm{s}$077 & --01$^\circ$09$^\prime$40$\farcs$42 & 0\farcs02\,$\times$\,0\farcs01 & 2017.1.01301.S & 26 & 43 & 0\farcs27$\times$0\farcs23 & 0.24 \\
P167--13 & 11$^\mathrm{h}$10$^\mathrm{m}$33$.\!\!^\mathrm{s}$963 & --13$^\circ$29$^\prime$45$\farcs$73 & 0\farcs07\,$\times$\,0\farcs04 & 2016.1.00544.S & 43 & 47 & 0\farcs33$\times$0\farcs22 & 0.20 \\
J1120+0641 & 11$^\mathrm{h}$20$^\mathrm{m}$01$.\!\!^\mathrm{s}$463 & +06$^\circ$41$^\prime$23$\farcs$79 & 0\farcs02\,$\times$\,0\farcs03 & 2012.1.00882.S & 161 & 38--47 & 0\farcs24$\times$0\farcs23 & 0.13 \\
P183+05 & 12$^\mathrm{h}$12$^\mathrm{m}$26$.\!\!^\mathrm{s}$969 & +05$^\circ$05$^\prime$33$\farcs$49 & 0\farcs05\,$\times$\,0\farcs02 & 2016.1.00544.S & 47 & 39 & 0\farcs27$\times$0\farcs24 & 0.32 \\
J1306+0356 & 13$^\mathrm{h}$06$^\mathrm{m}$08$.\!\!^\mathrm{s}$259 & +03$^\circ$56$^\prime$26$\farcs$19 & 0\farcs02\,$\times$\,0\farcs02 & 2017.1.01301.S & 28 & 43 & 0\farcs25$\times$0\farcs24 & 0.45 \\
J1319+0950 & 13$^\mathrm{h}$19$^\mathrm{m}$11$.\!\!^\mathrm{s}$291 & +09$^\circ$50$^\prime$51$\farcs$49 & 0\farcs02\,$\times$\,0\farcs01 & 2012.1.00240.S & 50 & 34 & 0\farcs27$\times$0\farcs21 & 0.29 \\
J1342+0928 & 13$^\mathrm{h}$42$^\mathrm{m}$08$.\!\!^\mathrm{s}$100 & +09$^\circ$28$^\prime$38$\farcs$36 & 0\farcs17\,$\times$\,0\farcs10 & 2017.1.00396.S & 114 & 43 & 0\farcs20$\times$0\farcs15 & 0.13 \\
P231--20 & 15$^\mathrm{h}$26$^\mathrm{m}$37$.\!\!^\mathrm{s}$844 & --20$^\circ$50$^\prime$00$\farcs$88 & 0\farcs08\,$\times$\,0\farcs08 & 2016.1.00544.S & 43 & 44 & 0\farcs21$\times$0\farcs12 & 0.21 \\
P308--21 & 20$^\mathrm{h}$32$^\mathrm{m}$09$.\!\!^\mathrm{s}$996 & --21$^\circ$14$^\prime$02$\farcs$38 & 0\farcs08\,$\times$\,0\farcs08 & 2016.A.00018.S & 56 & 44--45 & 0\farcs28$\times$0\farcs22 & 0.17 \\
J2054--0005 & 20$^\mathrm{h}$54$^\mathrm{m}$06$.\!\!^\mathrm{s}$496 & --00$^\circ$05$^\prime$14$\farcs$57 & 0\farcs10\,$\times$\,0\farcs10 & 2018.1.00908.S & 85 & 46 & 0\farcs15$\times$0\farcs11 & 0.28 \\
J2100--1715 & 21$^\mathrm{h}$00$^\mathrm{m}$54$.\!\!^\mathrm{s}$698 & --17$^\circ$15$^\prime$22$\farcs$00 & 0\farcs06\,$\times$\,0\farcs06 & 2017.1.01301.S & 25 & 45 & 0\farcs24$\times$0\farcs20 & 0.38 \\
P323+12 & 21$^\mathrm{h}$32$^\mathrm{m}$33$.\!\!^\mathrm{s}$178 & +12$^\circ$17$^\prime$55$\farcs$07 & 0\farcs02\,$\times$\,0\farcs05 & 2018.1.00908.S & 43 & 49 & 0\farcs12$\times$0\farcs09 & 0.29 \\
J2318--3113 & 23$^\mathrm{h}$18$^\mathrm{m}$18$.\!\!^\mathrm{s}$369 & --31$^\circ$13$^\prime$46$\farcs$42 & 0\farcs06\,$\times$\,0\farcs04 & 2017.1.01301.S & 24 & 45 & 0\farcs25$\times$0\farcs22 & 0.33 \\
J2318--3029 & 23$^\mathrm{h}$18$^\mathrm{m}$33$.\!\!^\mathrm{s}$099 & --30$^\circ$29$^\prime$33$\farcs$58 & 0\farcs04\,$\times$\,0\farcs03 & 2018.1.00908.S & 40 & 48 & 0\farcs10$\times$0\farcs10 & 0.20 \\
J2348--3054 & 23$^\mathrm{h}$48$^\mathrm{m}$33$.\!\!^\mathrm{s}$352 & --30$^\circ$54$^\prime$10$\farcs$35 & 0\farcs07\,$\times$\,0\farcs09 & 2015.1.00399.S & 52 & 44 & 0\farcs18$\times$0\farcs14 & 0.25 \\
P359--06 & 23$^\mathrm{h}$56$^\mathrm{m}$32$.\!\!^\mathrm{s}$452 & --06$^\circ$22$^\prime$59$\farcs$31 & 0\farcs08\,$\times$\,0\farcs04 & 2017.1.01301.S & 26 & 44 & 0\farcs25$\times$0\farcs21 & 0.27 \\
\enddata
\tablenotetext{a}{Optical/near-infrared quasar position and uncertainty (see Section~\ref{sec:astrometry} for details).}
\tablenotetext{b}{On-source integration time.}
\tablenotetext{c}{Number of antennas.}
\tablenotetext{d}{Beam size measured in the integrated \cii\ maps.}
\end{deluxetable*}

This paper is organized as follows. In Section~\ref{sec:observations}, we introduce the sample of high-redshift quasar host galaxies analyzed in this paper (Section~\ref{sec:sample}) and provide the details of the observations and reduction (Section~\ref{sec:obs}). In Section~\ref{sec:results} we present our main results, from the spectra (Section~\ref{sec:ciispectra}), the moment-zero maps (Section~\ref{sec:maps}), and from the search for line emitters in the field (Section~\ref{sec:lineemitters}). Our results are discussed in Section~\ref{sec:discussion}, followed by a summary in Section~\ref{sec:summary}.

We adopt a concordance cosmology with $\Omega_M=0.3$, $\Omega_\Lambda=0.7$, and $H_0=70$\,\kms\,Mpc$^{-1}$, which is consistent with the recent measurements of Planck \citep[][]{pla16}. In this cosmology at a redshift of 6.0 (7.0), 1\arcsec\ corresponds to 5.7 (5.2) kpc. Star-formation rates are computed using the \citet{kro03} initial mass function.

\section{Observations}
\label{sec:observations}

\subsection{Sample}
\label{sec:sample}

\begin{figure}
\begin{center}
\includegraphics[width=\columnwidth]{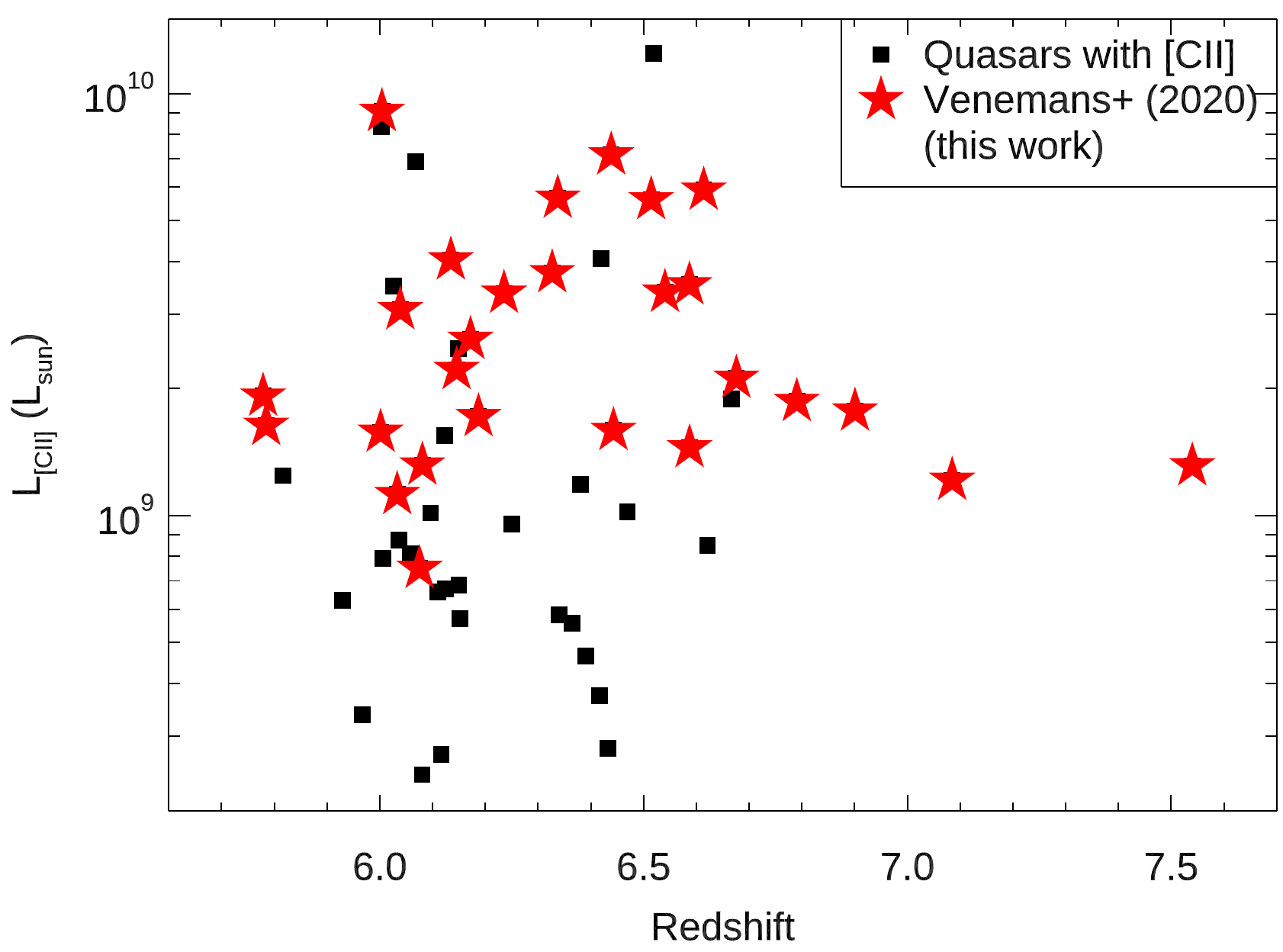}
\label{fig:lumsample}
\caption{\cii\ luminosity as a function of redshift for the quasars in our sample (red stars) and for all known $z>5.7$ quasars from the literature (black squares). The observations presented here targeted the brighter quasar hosts known at these redshifts. Our sample contains 47\% of all quasars at $z>5.7$ with a \cii\ detection and 68\% of those with \lcii\,$>10^9$\,\lsun.}
\end{center}
\end{figure}

The sample presented in this paper consists of all quasars at $z>5.7$ for which high spatial resolution (here defined as a spatial resolution better than $\sim$0\farcs35, or $\lesssim$2\,kpc) was obtained with ALMA up to the end of Cycle 6 (2019 September). In total, 27 quasars were found to have sufficiently high spatial resolution ALMA imaging. Note that this sample of $z\sim6$ quasars is different from the one presented in \citet{dec18} and \citet{ven18}, despite containing the same number of objects. The quasars and the details of the observations are listed in Table~\ref{tab:sample}. In Appendix~\ref{sec:notes} we describe some individual objects in more detail. The coordinates in the table represent the position of the optical/near-infrared point source (the active galactic nucleus); see Section~\ref{sec:astrometry} below for the details. All quasars in our sample have previously been detected in the redshifted \cii\ emission and far-infrared continuum, mostly in low-resolution imaging \citep[with beam sizes $\gtrsim$0\farcs7; e.g.,][]{dec18}. In Figure~\ref{fig:lumsample} we show the redshift and \cii\ luminosity distribution of our sample.

\subsection{ALMA Observations and Data Reduction}
\label{sec:obs}

The ALMA data presented in this paper were obtained as part of several projects observed between Cycle 1 and Cycle 6. The project IDs are 2012.1.00240.S (PI: R.\ Wang), 2012.1.00882.S (PI: B.\ Venemans), 2013.1.00273.S (PI: B.\ Venemans), 2015.1.00399.S (PI: B.\ Venemans), 2015.1.00692.S (PI: X.\ Fan), 2016.1.00544 (PI: E.\ Ba\~nados), 2016.A.00018.S (PI: R.\ Decarli), 2017.1.00396.S (PI: E.\ Ba\~nados), 2017.1.01301.S (PI: F.\ Walter), and 2018.1.00908.S (PI: F.\ Walter); see Table~\ref{tab:sample}. The observations were carried out between 2013 July 5 and 2019 September 24. The on-source integration time for each dataset ranged from 16 to 161\,min, and the spatial resolution varies from 0\farcs1 to 0\farcs33. The number of antennas was 34 to 49 (see Table~\ref{tab:sample} for details). The beam size listed in Table~\ref{tab:sample} is that in the \cii\ maps, created with {\sc robust=0.5} weighting (see below).

The setup of the observations differed slightly from source to source, but for most sources two overlapping sidebands (SPWs), of 1.875\,GHz wide each, were used to cover the redshifted \cii\ line (rest frequency of 1900.54\,GHz). The frequency coverage around the \cii\ line in these cases was about 3.3\,GHz, which corresponds to $\sim$3600\,\kms\ at $z=6$. Two other bandpasses of 1.875\,GHz each were placed $\sim$15\,GHz away, probing only continuum emission.

The data were reduced following the standard reduction steps using the Common Astronomy Software Applications package \citep[CASA;][]{mul07}. The calibration of the data was performed using the standard pipeline, with a few exceptions. For details on the data reduction and calibration, we refer to the accompanying paper by \citet{nov20}. The reduced measurement sets were subsequently imaged using the CASA task {\sc tclean} using Briggs weighting with the robust parameter set to {\sc robust=0.5}. This value optimizes both signal-to-noise ratio (S/N) and spatial resolution. For all cubes and maps, we cleaned down to 2$\sigma$ with  $\sigma$ being the rms noise. For each quasar, we created the several datasets produced by the following procedure:

\begin{enumerate}

\item We first created two data cubes for each quasar field with a channel width of 30\,MHz (corresponding to 32--40\,\kms, depending on the redshift of the source), one covering the two bandpasses around the \cii\ line and one covering the  bandpasses $\sim$15\,GHz away from the line. For J0109--3047 and J0305--3150 we created three data cubes as the continuum bandpasses were placed differently in Cycle 2 and Cycle 3 (Table~\ref{tab:sample}), resulting in a total number of 56 data cubes. The data cube containing the \cii\ emission line was subsequently used to extract the total spectrum of the quasar host galaxy (see Section~\ref{sec:ciispectra}). In all 56 data cubes, we searched for emission line galaxies in the field (see Section~\ref{sec:lineemitters}). The rms in the bandpasses containing the redshifted \cii\ line ranged from 0.13 to 0.54\,mJy\,beam$^{-1}$ per 30\,MHz channel (Table~\ref{sec:sample}).

\item We subtracted the continuum in $uv$ space using the task {\sc uvcontsub} with the polynomial order of the fit set to {\sc fitorder=1}. The continuum was defined as all channels in two bandpasses surrounding the \cii\ line that were at least 1.25\,$\times$\,FWHM of the \cii\ line away from the peak of the line. 

\item From the continuum-subtracted measurement sets, we created \cii\ data cubes to analyze the kinematics of the emission line (discussed in Neeleman M.\ et al\ 2020, in preparation).

\item \cii\ intensity maps were created by averaging the channels 1.2\,$\times$\,FWHM around the peak of the \cii\ line. This width results in the highest S/N \cii\ map if the emission line is Gaussian (and the noise is constant); see \citet{nov20}. The line flux in such a map is $\sim$84\% of the total flux. The maps are presented in Section~\ref{sec:maps}.

\item We created maps of the continuum emission by setting the deconvolver in {\sc tclean} to {\sc mtmfs} (multiterm multifrequency synthesis) with the number of Taylor coefficients set to two and the redshifted frequency of the \cii\ line as reference frequency. The continuum in this step was defined as all channels in all four bandpasses that were at least 1.25\,$\times$\,FWHM away from the peak of the \cii\ line. This procedure creates continuum maps at the frequency of the \cii\ emission line by fitting a second-order polynomial to the continuum channels. 

\begin{figure*}
\begin{center}
\includegraphics[width=\textwidth]{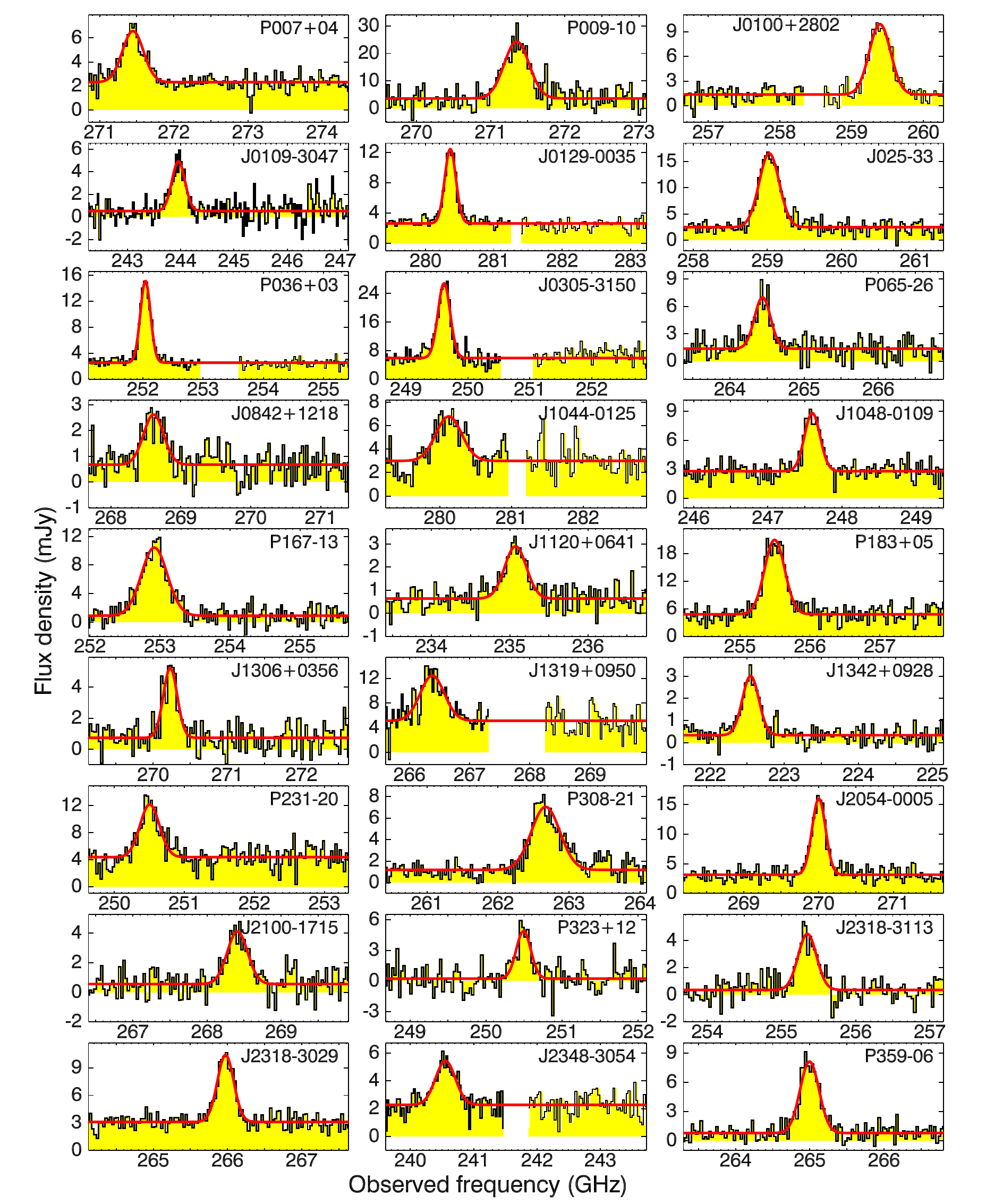}
\caption{\cii\ spectra of the 27 quasars in our sample, based on the two bandpasses encompassing the \cii\ emission line. The spectra were extracted using aperture photometry with residual scaling \citep[discussed in detail in][]{nov19,nov20}. The solid red line represents the Gaussian+continuum fit to the spectra (tabulated in Table~\ref{tab:results}). \label{fig:figwithspectra}
}
\end{center}
\end{figure*}

\end{enumerate}

\newpage
\subsection{Optical/Near-infrared Coordinates}
\label{sec:astrometry}

To compare the location of the accreting black hole (the quasar) to that of the host galaxy as traced by ALMA, we required accurate positions of the optical/near-infrared quasar on the same reference frame as the ALMA observations. The optical or near-infrared positions of the quasars were taken from catalogs of large surveys such as the Pan-STARRS1 survey \citep{cha16} and the near-infrared VIKING survey \citep{edg13}. We corrected the position of stars within 2\arcmin\ of the quasar in the optical/near-infrared catalogs that had a match in the Gaia DR2 catalog \citep{gaia18}. To compute the uncertainty of these astrometrically corrected positions, we calculated the standard deviation of the offsets between the corrected and Gaia position of stars close to the quasar. The resulting uncertainties in the astrometry range from 0\farcs01 to 0\farcs15. The optical/near-infrared positions and the uncertainties are listed in Table~\ref{tab:sample}. It should be noted that the position and uncertainties in the table do not account for systematic offsets that are due to differential chromatic refraction \citep[see, e.g.,][]{kac09}. Because the refractive index of air is dependent on the wavelength, the difference in color of the quasars with respect to those of the stars used for the astrometric calibration could result in a systematic positional offset, especially if a quasar has been observed at high airmass. We will discuss this effect in Section~\ref{sec:location}.

\section{Results}
\label{sec:results}

\subsection{\cii\ Spectra}
\label{sec:ciispectra}

To obtain a total spectrum of each quasar host galaxy, we extracted spectra in the \cii+continuum data cubes. As our sources are all resolved at the spatial resolution of our observations, the emission could be extended over multiple beams. To obtain a full spectrum of a resolved source, we extracted spectra in an aperture while taking both clean and uncleaned (residual) components into account by applying the residual scale method. This method is described in detail in \citet[][]{nov19,nov20}. The apertures used for the extraction were chosen manually after visual inspection and had radii between 0\farcs4 and 1\farcs5, depending on the extent of the source (see Figure~\ref{fig:allmaps}). The spectra for our 27 sources are shown in Figure~\ref{fig:figwithspectra}. 

We fitted a single Gaussian with a constant continuum to each of the extracted spectra. The fits are shown with red lines in Figure~\ref{fig:figwithspectra}. The best-fit parameters are subsequently used to derive the properties of the \cii\ line and FIR continuum in the quasar host. The measured \cii\ redshift, line flux, line width, equivalent width, and continuum flux density are listed in Table~\ref{tab:results}. The equivalent width of the \cii\ emission was computed via

\begin{equation}
\mathrm{EW}_\mathrm{[CII]}/\mu\mathrm{m} = 1000\,\frac{F_\mathrm{[CII]}}{S_{1900\,\mathrm{GHz}}}\,\frac{\lambda_\mathrm{[CII],0}}{c},
\end{equation}

\noindent
with $F_\mathrm{[CII]}$ being the observed \cii\ flux in Jy\,\kms, $S_{1900\,\mathrm{GHz}}$ the observed continuum flux density at a rest-frame frequency of 1900\,GHz in mJy, and $\lambda_\mathrm{[CII],0}$ the rest wavelength of the \cii\ emission line of $\lambda_\mathrm{[CII],0}=157.74\,\mu$m. The \cii\ luminosities were derived from the measured line fluxes using

\begin{equation}
L_\mathrm{[CII]} / L_\odot = 1.04\times10^{-3}\,\nu_\mathrm{[CII],obs}\,F_\mathrm{[CII]}\,D_L^2,
\end{equation}

\noindent
with $\nu_\mathrm{[CII],obs}$ being the observed frequency of the \cii\ line in GHz, and $D_L$ the luminosity distance in Mpc. 

To estimate the far-infrared luminosity, \lfir, of the quasar host galaxies, we need to assume a function form of the dust spectral energy distribution (SED) and integrate the emission between the rest-frame wavelengths of 42.5 and 122.5\,$\mu$m \citep[e.g.,][]{hel88}. Following the literature, we assume that the dust SED can be described by a modified blackbody with a dust temperature of $T_d=47$\,K and an emissivity index of $\beta=1.6$ \citep[e.g.,][]{bee06,lei14,ven18}. Under these assumptions, the FIR luminosity can be derived from the continuum flux density using the following equation:

\begin{equation}
\label{eq:fir}
L_\mathrm{FIR} / L_\odot = 3.86\times10^3\,\frac{S_{1900\,\mathrm{GHz}}}{f_\mathrm{CMB}(z)}\,\frac{D_L^2}{1+z},
\end{equation}

\noindent
with $f_\mathrm{CMB}$ being the fraction of the continuum flux density measured against the cosmic microwave background \citep[CMB; see][for a detailed review of this effect]{dac13}, given by

\begin{equation}
f_\mathrm{CMB} = 1 - \frac{B_{\nu,\mathrm{[CII]},0}(T_\mathrm{CMB}(z)}{B_{\nu,\mathrm{[CII]},0}(T_d=47\,K))},
\end{equation}

\noindent
with $B_{\nu,\mathrm{[CII]},0}$ being the Planck function at the rest frequency of the \cii\ line ($\nu_\mathrm{[CII],0} = 1900.54$\,GHz), and $T_\mathrm{CMB}(z)$ the temperature of the CMB at redshift $z$. We here ignore the dust heating by the CMB, which has a negligible effect on the derived FIR luminosity ($<$1\%) for our assumed dust temperature. The systematic uncertainty on the FIR luminosity due to the unknown shape of the dust SED is a factor of $\sim$2--3 \citep[see][for an extended discussion]{ven18}.

\begin{deluxetable*}{lcccccccc}[t]
\tablecaption{Results from Spectral Fitting of the Total \cii\ Spectra Shown in Figure~\ref{fig:figwithspectra}. \label{tab:results}}
\tablewidth{0pt}
\tablehead{\colhead{Name} & \colhead{$z_\mathrm{[CII]}$} &
\colhead{$F_\mathrm{[CII]}$} & \colhead{FWHM$_\mathrm{[CII]}$} &
\colhead{$S_\mathrm{1900\,GHz}$} & \colhead{EW$_\mathrm{[CII]}$} &
\colhead{\lcii} & \colhead{\lfir} & \colhead{\lcii/\lfir} \\
\colhead{} & \colhead{} & \colhead{(Jy\,\kms)} &
\colhead{(\kms)} & \colhead{(mJy)} & \colhead{($\mu$m)} &
\colhead{($10^9$\,\lsun)} & \colhead{($10^{12}$\,\lsun)} & \colhead{($10^{-3}$)}}
\startdata
P007+04 & 6.0015$\pm$0.0002 & 1.67$\pm$0.10 & 370$\pm$22 & 2.33$\pm$0.06 & 0.38$\pm$0.02 & 1.58$\pm$0.09 & 4.52$\pm$0.11 & 0.35$\pm$0.02 \\
P009--10 & 6.0040$\pm$0.0003 & 9.60$\pm$0.70 & 437$\pm$33 & 3.66$\pm$0.36 & 1.38$\pm$0.17 & 9.07$\pm$0.66 & 7.13$\pm$0.69 & 1.27$\pm$0.15 \\
J0100+2802 & 6.3269$\pm$0.0002 & 3.69$\pm$0.17 & 405$\pm$20 & 1.37$\pm$0.09 & 1.42$\pm$0.11 & 3.76$\pm$0.17 & 2.92$\pm$0.18 & 1.29$\pm$0.10 \\
J0109--3047 & 6.7904$\pm$0.0003 & 1.65$\pm$0.14 & 354$\pm$34 & 0.52$\pm$0.08 & 1.67$\pm$0.29 & 1.87$\pm$0.16 & 1.26$\pm$0.19 & 1.49$\pm$0.26 \\
J0129--0035 & 5.7788$\pm$0.0001 & 2.15$\pm$0.08 & 206$\pm$9 & 2.61$\pm$0.06 & 0.43$\pm$0.02 & 1.92$\pm$0.07 & 4.76$\pm$0.11 & 0.40$\pm$0.02 \\
J025--33 & 6.3373$\pm$0.0002 & 5.53$\pm$0.21 & 370$\pm$16 & 2.49$\pm$0.11 & 1.17$\pm$0.07 & 5.65$\pm$0.22 & 5.31$\pm$0.24 & 1.06$\pm$0.06 \\
P036+03 & 6.5405$\pm$0.0001 & 3.16$\pm$0.09 & 237$\pm$7 & 2.55$\pm$0.05 & 0.65$\pm$0.02 & 3.38$\pm$0.09 & 5.77$\pm$0.12 & 0.59$\pm$0.02 \\
J0305--3150\tablenotemark{a} & 6.6139$\pm$0.0002 & 5.43$\pm$0.33 & 225$\pm$15 & 5.34$\pm$0.19 & 0.53$\pm$0.04 & 5.90$\pm$0.36 & 12.30$\pm$0.44 & 0.48$\pm$0.03 \\
P065--26 & 6.1871$\pm$0.0003 & 1.74$\pm$0.17 & 289$\pm$31 & 1.37$\pm$0.11 & 0.67$\pm$0.09 & 1.71$\pm$0.17 & 2.80$\pm$0.23 & 0.61$\pm$0.08 \\
J0842+1218 & 6.0754$\pm$0.0005 & 0.78$\pm$0.10 & 378$\pm$52 & 0.68$\pm$0.06 & 0.61$\pm$0.09 & 0.75$\pm$0.10 & 1.34$\pm$0.11 & 0.56$\pm$0.09 \\
J1044--0125 & 5.7846$\pm$0.0005 & 1.83$\pm$0.23 & 454$\pm$60 & 3.00$\pm$0.12 & 0.32$\pm$0.04 & 1.64$\pm$0.21 & 5.48$\pm$0.22 & 0.30$\pm$0.04 \\
J1048--0109 & 6.6759$\pm$0.0002 & 1.92$\pm$0.14 & 299$\pm$24 & 2.79$\pm$0.08 & 0.36$\pm$0.03 & 2.11$\pm$0.15 & 6.54$\pm$0.19 & 0.32$\pm$0.03 \\
P167--13 & 6.5144$\pm$0.0003 & 5.27$\pm$0.25 & 519$\pm$25 & 0.89$\pm$0.12 & 3.11$\pm$0.44 & 5.60$\pm$0.27 & 2.00$\pm$0.27 & 2.80$\pm$0.40 \\
J1120+0641 & 7.0848$\pm$0.0004 & 1.01$\pm$0.09 & 416$\pm$39 & 0.64$\pm$0.05 & 0.83$\pm$0.10 & 1.21$\pm$0.11 & 1.66$\pm$0.12 & 0.73$\pm$0.08 \\
P183+05 & 6.4386$\pm$0.0002 & 6.84$\pm$0.31 & 397$\pm$19 & 4.79$\pm$0.16 & 0.75$\pm$0.04 & 7.15$\pm$0.32 & 10.53$\pm$0.36 & 0.68$\pm$0.04 \\
J1306+0356 & 6.0330$\pm$0.0002 & 1.18$\pm$0.11 & 246$\pm$26 & 0.74$\pm$0.08 & 0.83$\pm$0.12 & 1.12$\pm$0.11 & 1.46$\pm$0.16 & 0.77$\pm$0.11 \\
J1319+0950 & 6.1347$\pm$0.0005 & 4.14$\pm$0.43 & 532$\pm$57 & 5.13$\pm$0.22 & 0.42$\pm$0.05 & 4.03$\pm$0.42 & 10.36$\pm$0.44 & 0.39$\pm$0.04 \\
J1342+0928 & 7.5400$\pm$0.0003 & 1.00$\pm$0.07 & 353$\pm$27 & 0.34$\pm$0.04 & 1.57$\pm$0.20 & 1.32$\pm$0.09 & 0.99$\pm$0.10 & 1.34$\pm$0.17 \\
P231--20 & 6.5869$\pm$0.0004 & 3.26$\pm$0.28 & 393$\pm$35 & 4.37$\pm$0.15 & 0.39$\pm$0.04 & 3.53$\pm$0.30 & 9.99$\pm$0.34 & 0.35$\pm$0.03 \\
P308--21 & 6.2355$\pm$0.0003 & 3.37$\pm$0.19 & 541$\pm$32 & 1.18$\pm$0.08 & 1.50$\pm$0.13 & 3.37$\pm$0.19 & 2.45$\pm$0.16 & 1.37$\pm$0.12 \\
J2054--0005 & 6.0389$\pm$0.0001 & 3.23$\pm$0.14 & 236$\pm$12 & 3.15$\pm$0.10 & 0.54$\pm$0.03 & 3.08$\pm$0.14 & 6.20$\pm$0.19 & 0.50$\pm$0.03 \\
J2100--1715 & 6.0807$\pm$0.0004 & 1.37$\pm$0.14 & 361$\pm$41 & 0.56$\pm$0.08 & 1.27$\pm$0.22 & 1.31$\pm$0.14 & 1.12$\pm$0.16 & 1.17$\pm$0.20 \\
P323+12 & 6.5872$\pm$0.0004 & 1.34$\pm$0.17 & 271$\pm$38 & 0.23$\pm$0.12 & 3.04$\pm$1.58 & 1.45$\pm$0.19 & 0.53$\pm$0.27 & 2.74$\pm$1.42 \\
J2318--3113 & 6.4429$\pm$0.0003 & 1.52$\pm$0.14 & 344$\pm$34 & 0.36$\pm$0.08 & 2.21$\pm$0.50 & 1.59$\pm$0.14 & 0.79$\pm$0.17 & 2.00$\pm$0.46 \\
J2318--3029 & 6.1456$\pm$0.0002 & 2.27$\pm$0.12 & 293$\pm$17 & 3.11$\pm$0.07 & 0.38$\pm$0.02 & 2.22$\pm$0.12 & 6.29$\pm$0.14 & 0.35$\pm$0.02 \\
J2348--3054 & 6.9007$\pm$0.0005 & 1.53$\pm$0.16 & 457$\pm$49 & 2.28$\pm$0.07 & 0.35$\pm$0.04 & 1.77$\pm$0.18 & 5.66$\pm$0.18 & 0.31$\pm$0.03 \\
P359--06 & 6.1719$\pm$0.0002 & 2.66$\pm$0.13 & 341$\pm$18 & 0.79$\pm$0.07 & 1.78$\pm$0.19 & 2.62$\pm$0.13 & 1.60$\pm$0.15 & 1.63$\pm$0.17 \\
\enddata
\tablenotetext{a}{Values taken from \citet{ven19}.}
\end{deluxetable*}

\begin{figure*}
\begin{center}
\includegraphics[width=\textwidth]{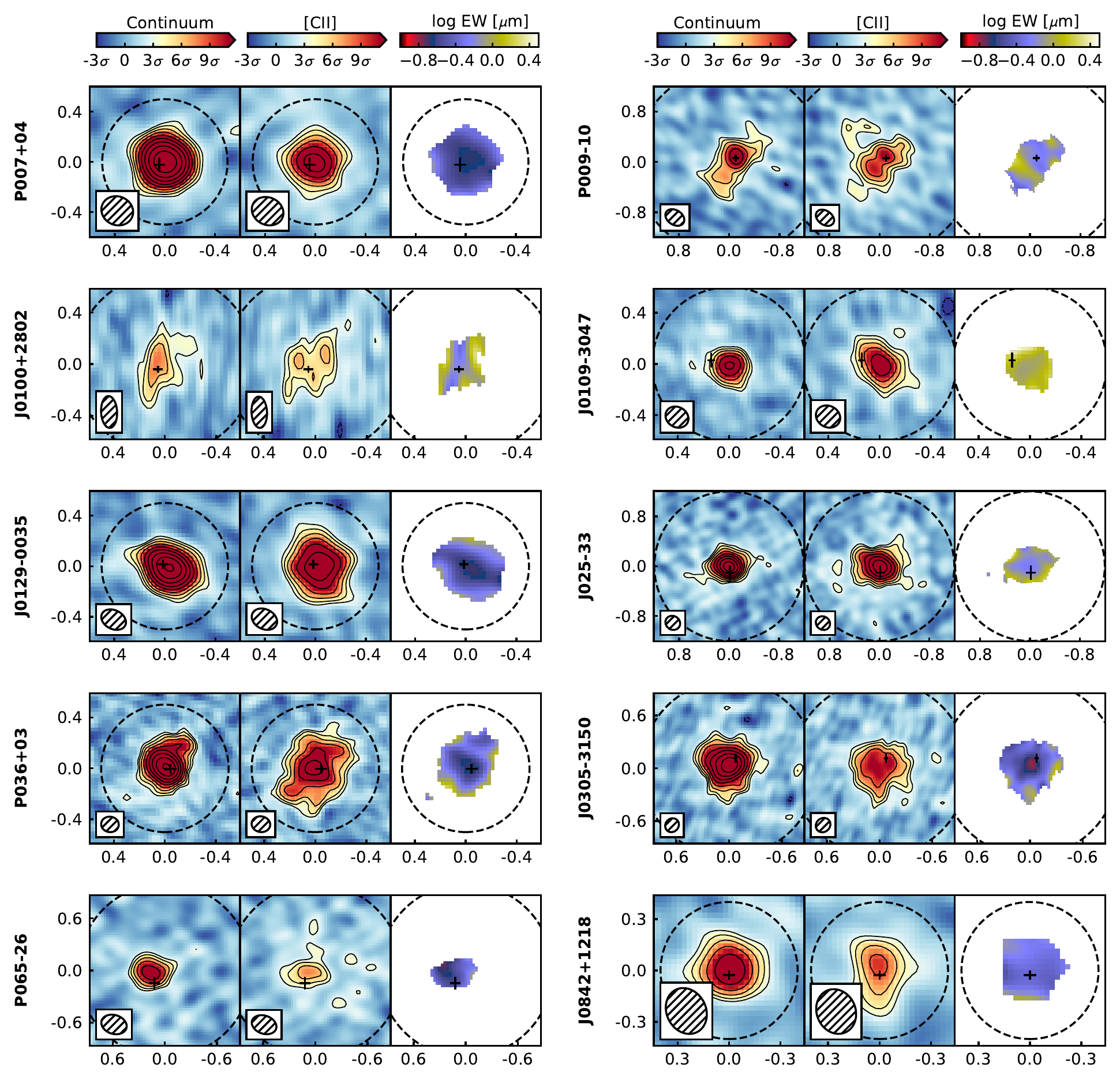}
\caption{Maps of the continuum, \cii, and \cii\ rest-frame equivalent width for the 27 quasars in our sample. The axes are in arcsec. The contours start at 3$\sigma$ and increase in powers of $\sqrt{2}$. The cross indicates the position of the optical/near-infrared quasar (Table~\ref{tab:sample}). The dashed line represents the aperture in which the total \cii\ spectrum was extracted, as shown in Figure~\ref{fig:figwithspectra}.}
\end{center}
\end{figure*}

\begin{figure*}\ContinuedFloat
\begin{center}
\includegraphics[width=\textwidth]{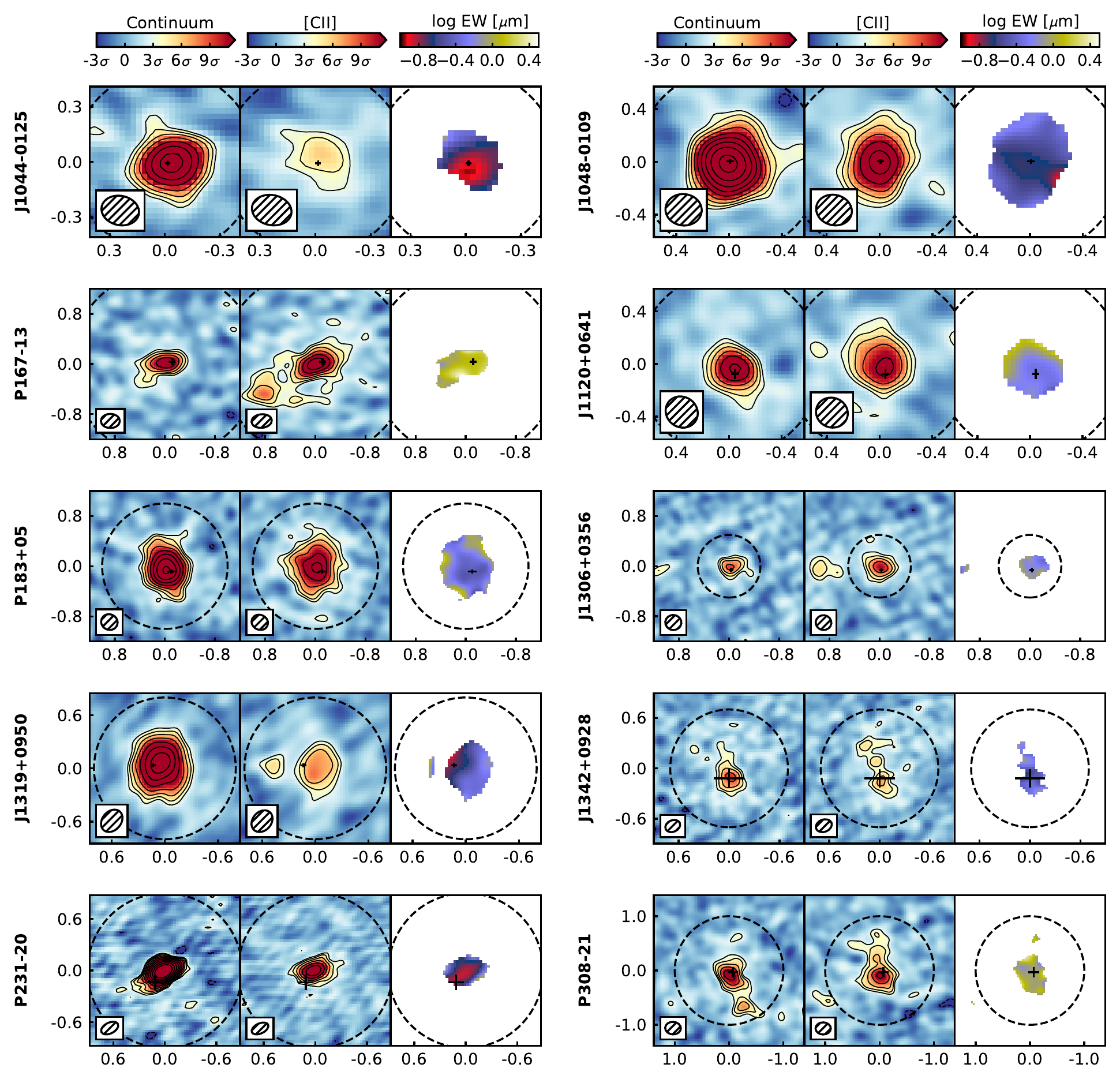}
\caption{Continued.}
\end{center}
\end{figure*}

\begin{figure*}\ContinuedFloat
\begin{center}
\includegraphics[width=\textwidth]{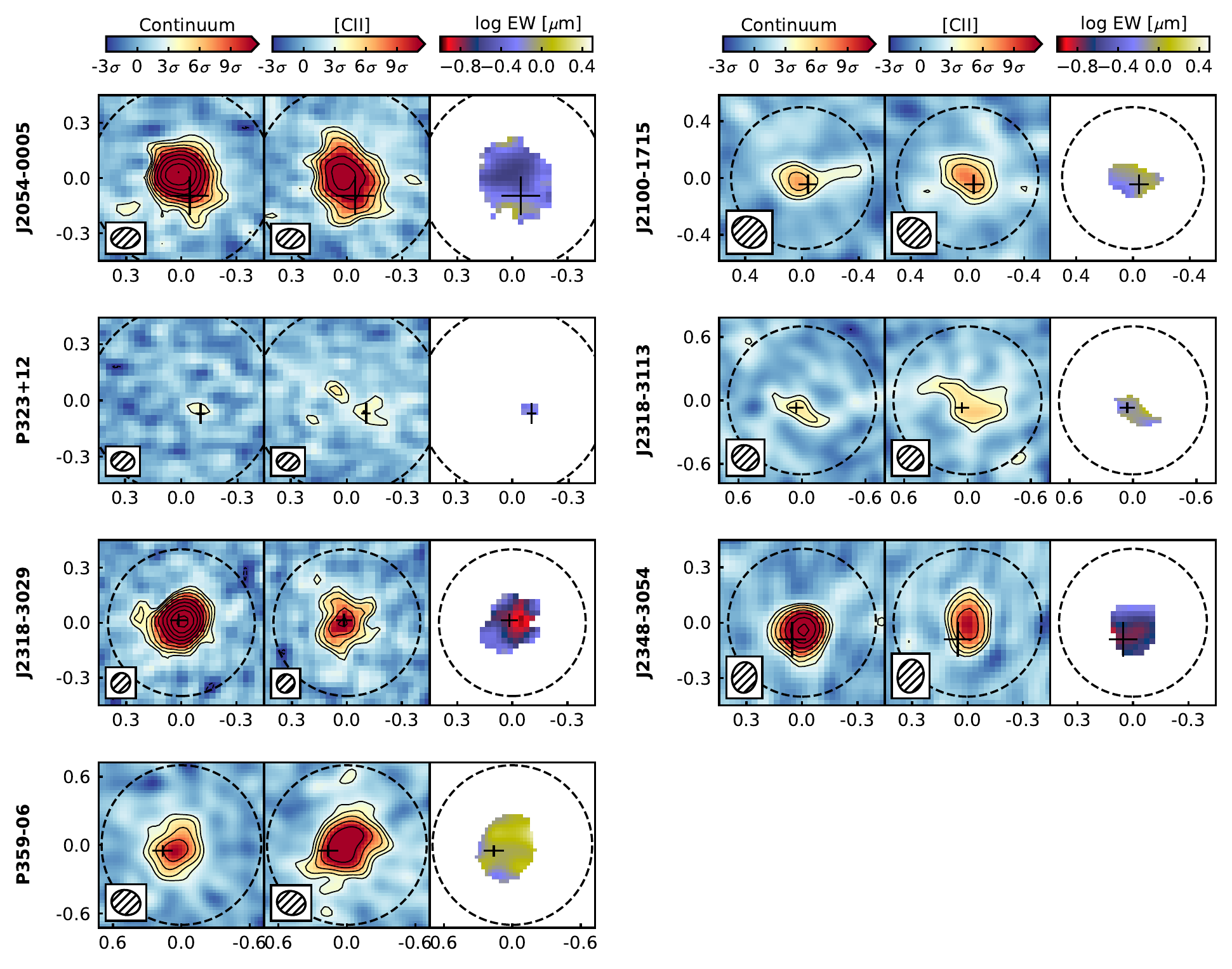}
\caption{Continued.}
\label{fig:allmaps}
\end{center}
\end{figure*}

Under the assumption that the dust is heated by star formation, we integrate the dust spectral energy distribution between the rest-frame wavelengths of 8 and 1000\,$\mu$m \citep[e.g.,][]{ken12} to obtain the total infrared luminosity \ltir\ and convert \ltir\ to an SFR. Adopting a modified blackbody with the canonical $T_d=47$\,K and $\beta=1.6$ to describe the dust SED, we find the correspondence between \lfir\ and \ltir\ is \ltir\,=\,1.41\,$\times$\,\lfir. If we subsequently apply the local relation between \ltir\ and SFR from \citet{mur11}, we compute the SFR from \lfir\ using

\begin{equation}
\label{eq:sfr}
\mathrm{SFR}\,/\,M_\odot\,\mathrm{yr}^{-1} = 2.1\times10^{-10}\,L_\mathrm{FIR}.
\end{equation}

The \cii\ luminosities of the quasar hosts in our sample range from \lcii\,$=7.5\times10^8$\,\lsun\ to \lcii\,$=9.1\times10^9$\,\lsun, with a mean of $3.0\times10^9$\,\lsun. This is slightly brighter than the average \cii\ line luminosity of all $z\sim6$ quasars studied so far ($2.4\times10^9$\,\lsun; see Figure~\ref{fig:lumsample}).

\subsection{2D Maps and Size Measurements}
\label{sec:maps}

\begin{deluxetable*}{lccccccc}
\tablecaption{\cii\ and Continuum Size Measurements from {\sc imfit}. \label{tab:size}}
\tablewidth{0pt}
\tablehead{\colhead{Name} & \colhead{\cii\ Size} &
\colhead{\cii\ Decl.\ Size} & \colhead{\cii\ Dec.\ Size} & \colhead{Cont.\ Size} &
\colhead{Cont.\ Dec.\ Size} & \colhead{Cont.\ Dec.\ Size} \\ 
\colhead{} & \colhead{} &  \colhead{} &
\colhead{(kpc$^2$)} & \colhead{} & \colhead{} & \colhead{(kpc$^2$)}} 
\startdata
P007+04 & 0\farcs33$\times$0\farcs32 & 0\farcs24$\times$0\farcs21 & 1.4$\times$1.2 & 0\farcs28$\times$0\farcs26 & 0\farcs13$\times$0\farcs11 & 0.7$\times$0.6 \\ 
P009--10 & 0\farcs87$\times$0\farcs68 & 0\farcs84$\times$0\farcs60 & 4.8$\times$3.4 & 0\farcs84$\times$0\farcs48 & 0\farcs80$\times$0\farcs35 & 4.6$\times$2.0 \\ 
J0100+2802 & 0\farcs54$\times$0\farcs42 & 0\farcs52$\times$0\farcs36 & 2.9$\times$2.0 & 0\farcs51$\times$0\farcs30 & 0\farcs47$\times$0\farcs24 & 2.6$\times$1.3 \\ 
J0109--3047 & 0\farcs35$\times$0\farcs26 & 0\farcs29$\times$0\farcs19 & 1.6$\times$1.0 & 0\farcs24$\times$0\farcs21 & 0\farcs15$\times$0\farcs13 & 0.8$\times$0.7 \\ 
J0129--0035 & 0\farcs37$\times$0\farcs33 & 0\farcs32$\times$0\farcs27 & 1.9$\times$1.6 & 0\farcs29$\times$0\farcs22 & 0\farcs20$\times$0\farcs16 & 1.2$\times$1.0 \\ 
J025--33 & 0\farcs50$\times$0\farcs38 & 0\farcs44$\times$0\farcs32 & 2.5$\times$1.7 & 0\farcs36$\times$0\farcs30 & 0\farcs27$\times$0\farcs19 & 1.5$\times$1.1 \\ 
P036+03 & 0\farcs45$\times$0\farcs32 & 0\farcs43$\times$0\farcs29 & 2.4$\times$1.6 & 0\farcs23$\times$0\farcs21 & 0\farcs19$\times$0\farcs16 & 1.0$\times$0.8 \\ 
J0305--3150 & 0\farcs54$\times$0\farcs51 & 0\farcs52$\times$0\farcs48 & 2.8$\times$2.6 & 0\farcs36$\times$0\farcs32 & 0\farcs32$\times$0\farcs28 & 1.7$\times$1.5 \\ 
P065--26 & 0\farcs90$\times$0\farcs73 & 0\farcs85$\times$0\farcs70 & 4.8$\times$3.9 & 0\farcs36$\times$0\farcs29 & 0\farcs21$\times$0\farcs18 & 1.2$\times$1.0 \\ 
J0842+1218 & 0\farcs40$\times$0\farcs33 & 0\farcs32$\times$0\farcs20 & 1.8$\times$1.1 & 0\farcs31$\times$0\farcs27 & $<$0\farcs28$\times$0\farcs23\,\, & $<$1.6$\times$1.3 \\ 
J1044--0125 & 0\farcs43$\times$0\farcs32 & 0\farcs37$\times$0\farcs27 & 2.2$\times$1.6 & 0\farcs27$\times$0\farcs22 & 0\farcs18$\times$0\farcs14 & 1.1$\times$0.8 \\ 
J1048--0109 & 0\farcs43$\times$0\farcs36 & 0\farcs37$\times$0\farcs24 & 2.0$\times$1.3 & 0\farcs33$\times$0\farcs33 & 0\farcs23$\times$0\farcs19 & 1.2$\times$1.0 \\ 
P167--13 & 0\farcs85$\times$0\farcs44 & 0\farcs79$\times$0\farcs37 & 4.3$\times$2.0 & 0\farcs57$\times$0\farcs32 & 0\farcs48$\times$0\farcs22 & 2.6$\times$1.2 \\ 
J1120+0641 & 0\farcs41$\times$0\farcs38 & 0\farcs33$\times$0\farcs30 & 1.7$\times$1.5 & 0\farcs33$\times$0\farcs26 & 0\farcs21$\times$0\farcs11 & 1.1$\times$0.6 \\ 
P183+05 & 0\farcs71$\times$0\farcs61 & 0\farcs66$\times$0\farcs55 & 3.6$\times$3.0 & 0\farcs51$\times$0\farcs43 & 0\farcs45$\times$0\farcs35 & 2.4$\times$1.9 \\ 
J1306+0356 & 0\farcs49$\times$0\farcs38 & 0\farcs42$\times$0\farcs29 & 2.4$\times$1.7 & 0\farcs38$\times$0\farcs30 & 0\farcs28$\times$0\farcs18 & 1.6$\times$1.0 \\ 
J1319+0950 & 0\farcs57$\times$0\farcs49 & 0\farcs52$\times$0\farcs41 & 3.0$\times$2.3 & 0\farcs47$\times$0\farcs41 & 0\farcs39$\times$0\farcs33 & 2.2$\times$1.8 \\ 
J1342+0928 & 0\farcs81$\times$0\farcs45 & 0\farcs80$\times$0\farcs40 & 4.0$\times$2.0 & 0\farcs53$\times$0\farcs31 & 0\farcs51$\times$0\farcs25 & 2.5$\times$1.2 \\ 
P231--20 & 0\farcs31$\times$0\farcs20 & 0\farcs24$\times$0\farcs16 & 1.3$\times$0.8 & 0\farcs24$\times$0\farcs15 & 0\farcs11$\times$0\farcs09 & 0.6$\times$0.5 \\ 
P308--21 & 0\farcs77$\times$0\farcs52 & 0\farcs73$\times$0\farcs44 & 4.1$\times$2.4 & 0\farcs99$\times$0\farcs44 & 0\farcs96$\times$0\farcs34 & 5.4$\times$1.9 \\ 
J2054--0005 & 0\farcs32$\times$0\farcs24 & 0\farcs29$\times$0\farcs19 & 1.7$\times$1.1 & 0\farcs20$\times$0\farcs19 & 0\farcs16$\times$0\farcs13 & 0.9$\times$0.7 \\ 
J2100--1715 & 0\farcs43$\times$0\farcs33 & 0\farcs36$\times$0\farcs25 & 2.0$\times$1.4 & 0\farcs45$\times$0\farcs21 & $<$0\farcs25$\times$0\farcs21\,\, & $<$1.4$\times$1.2 \\ 
P323+12 & 0\farcs61$\times$0\farcs26 & 0\farcs59$\times$0\farcs24 & 3.2$\times$1.3 & 0\farcs15$\times$0\farcs10 & $<$0\farcs12$\times$0\farcs10\,\, & $<$0.7$\times$0.5 \\ 
J2318--3113 & 0\farcs91$\times$0\farcs48 & 0\farcs87$\times$0\farcs42 & 4.8$\times$2.3 & 1\farcs04$\times$0\farcs33 & 1\farcs00$\times$0\farcs22 & 5.5$\times$1.2 \\
J2318--3029 & 0\farcs27$\times$0\farcs22 & 0\farcs25$\times$0\farcs20 & 1.4$\times$1.1 & 0\farcs18$\times$0\farcs15 & 0\farcs14$\times$0\farcs12 & 0.8$\times$0.7 \\ 
J2348--3054 & 0\farcs26$\times$0\farcs18 & 0\farcs18$\times$0\farcs10 & 1.0$\times$0.5 & 0\farcs19$\times$0\farcs18 & 0\farcs13$\times$0\farcs07 & 0.7$\times$0.4 \\ 
P359--06 & 0\farcs52$\times$0\farcs39 & 0\farcs47$\times$0\farcs30 & 2.6$\times$1.7 & 0\farcs45$\times$0\farcs38 & 0\farcs40$\times$0\farcs28 & 2.2$\times$1.6 \\ 
\enddata
\end{deluxetable*}

In Figure~\ref{fig:allmaps} we show the map of the continuum emission, the continuum-subtracted \cii\ map (created by averaging over 1.2$\times$FWHM$_\mathrm{[CII]}$ around the redshifted \cii\ line), and the EW$_\mathrm{[CII]}$ map.  
The \cii/FIR luminosity ratio will be discussed in Section~\ref{sec:ciifir}.

We measured the extent of the sources in both the \cii\ and continuum map by fitting a 2D Gaussian to the sources using the CASA task {\sc imfit}. The results are listed in Table~\ref{tab:size}. All of the quasars are resolved in our \cii\ maps, with deconvolved sizes ranging from 0\farcs18$\times$0\farcs10 from the most compact quasar host galaxy to 0\farcs85$\times$0\farcs70 for the most extended quasar host. In physical units, these sizes correspond to $\sim$1.0--4.8\,kpc. We stress that these size measurements only concern the bright, central emission and do not take extended low-surface-brightness emission into account. The extended emission will be described in an accompanying paper \citep{nov20}. In the continuum maps, three quasars (J0842+1218, J2100--1715, and P323+12) are not resolved as their measured sizes are consistent with the beam size. 

\begin{figure}
\begin{center}
\includegraphics[width=\columnwidth]{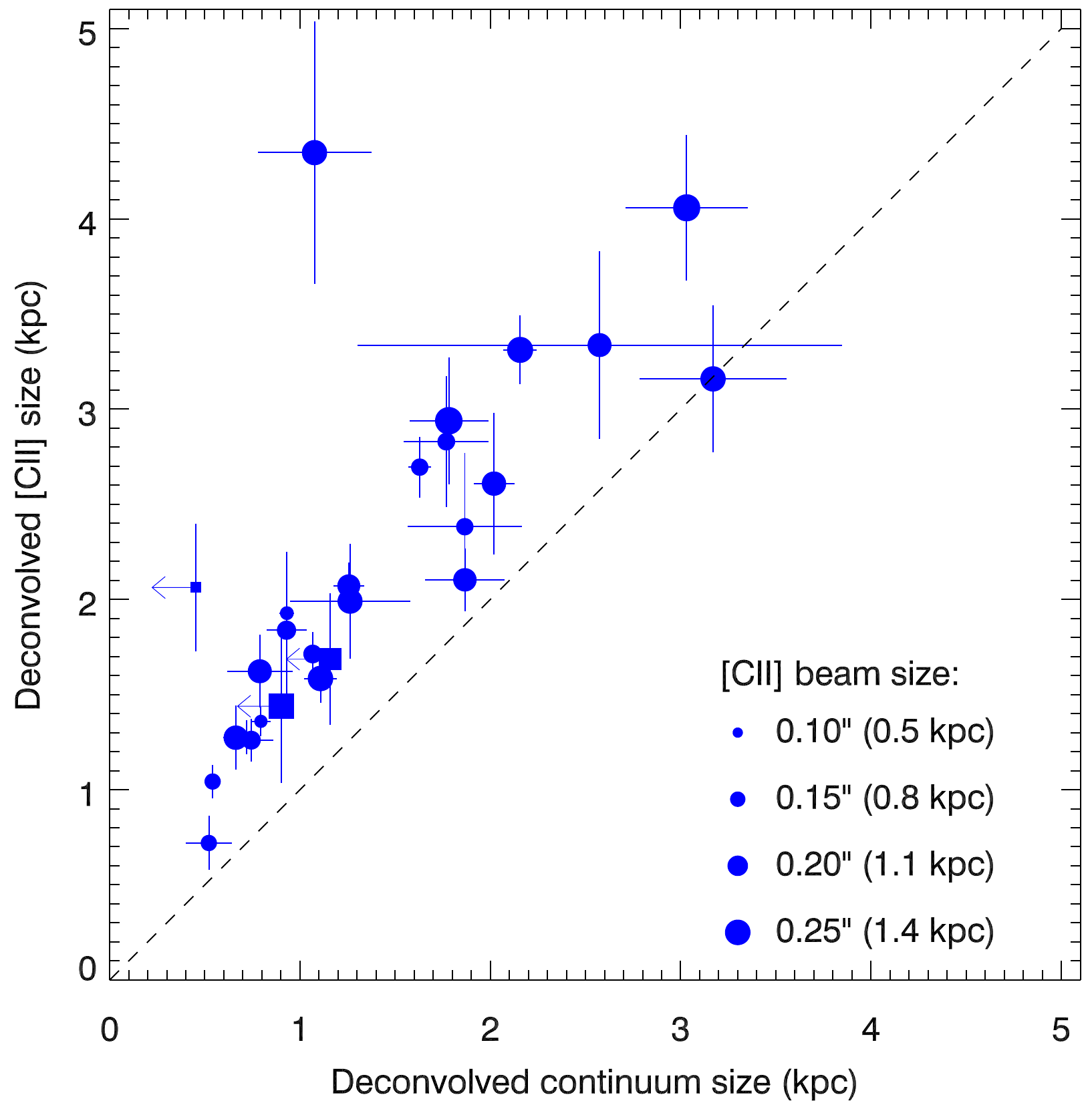}
\caption{Width in kpc of the 2D Gaussian fit to the quasar host galaxies in the continuum ($x$-axis) and \cii\ maps ($y$-axis). The sizes are deconvolved with the beam. The size of the symbols scales with the beam of the observations (listed in Table~\ref{tab:sample}). Quasar hosts unresolved in the continuum image are indicated with a square and an arrow. The dashed line represents the 1:1 line. In general, the continuum emission is more centrally concentrated than the \cii\ emission.}
\label{fig:sizes}
\end{center}
\end{figure}

The sizes of the \cii\ and continuum emission are shown in Figure~\ref{fig:sizes}. As reported previously for quasar hosts \citep[e.g.,][]{wan13,wil13,ven16} and for high-redshift star-forming galaxies \citep[e.g.,][]{cap15,gul18,ryb19,ryb20}, the continuum emission is more centrally concentrated than the \cii\ emission. As already shown in \citet{ven19} and further discussed in \citet{nov20}, this does not mean that the \cii\ emission is more extended than the continuum emission when considering faint emission: the size over which both \cii\ and continuum emission can be traced is approximately equal \citep[see][]{nov20}. The different radial profile of the line and continuum emission has an effect on the \cii/FIR luminosity ratio, which we discuss in Section~\ref{sec:ciifir}. 

The quasar host galaxies have a range of different morphologies. Although some of the sources show regular morphologies, others clearly show substructure. Six of the quasars have already-published companion galaxies detected in \cii. Most notably, \citet{dec17} reported bright \cii-emitting galaxies near four $z\sim6$ quasars: J0842+1218 and J2100--1715 have a companion galaxy several arcseconds (up to 50\,kpc) away from the quasar, while P231--20 has a more nearby companion galaxy and P308--21 is merging with a satellite galaxy \citep[see also][]{dec19}. \citet{wil17} reported a neighboring galaxy $\sim$1\arcsec\ away from quasar P167--13. This companion was confirmed by \citet{nee19} using higher-spatial-resolution observations. They also presented a new companion to J1306+0356. \citet{ven19} reported three \cii\ emitters in the field of J0305--3150, of which two were within 1\arcsec\ of the quasar. Besides P308--21, J1342+0942 also has a morphology consistent with undergoing a merger \citep{ban19}. 

We visually inspected all maps for nearby galaxies or disturbed \cii\ morphologies. We also performed an automated search for companion galaxies in the field (see Section~\ref{sec:lineemitters}). From the maps, we identify a close companion to the quasar J1319+0950, which has not been reported before. Despite the fact that caution is warranted when identifying substructure in low- or even moderate-S/N interferometric data \citep[e.g.,][]{hod16}, we found several sources, in particular P009--10, P065--26, and J2318--3113, that seem to have an irregular \cii\ morphology (see also Neeleman M.\ et al.\ 2020, in preparation). This indicates that a significant fraction of the $z\sim6$ quasar host galaxies are interacting with close companions or have disturbed morphologies.

\begin{figure*}
\begin{center}
\begin{tabular}{cc}
\includegraphics[width=\columnwidth]{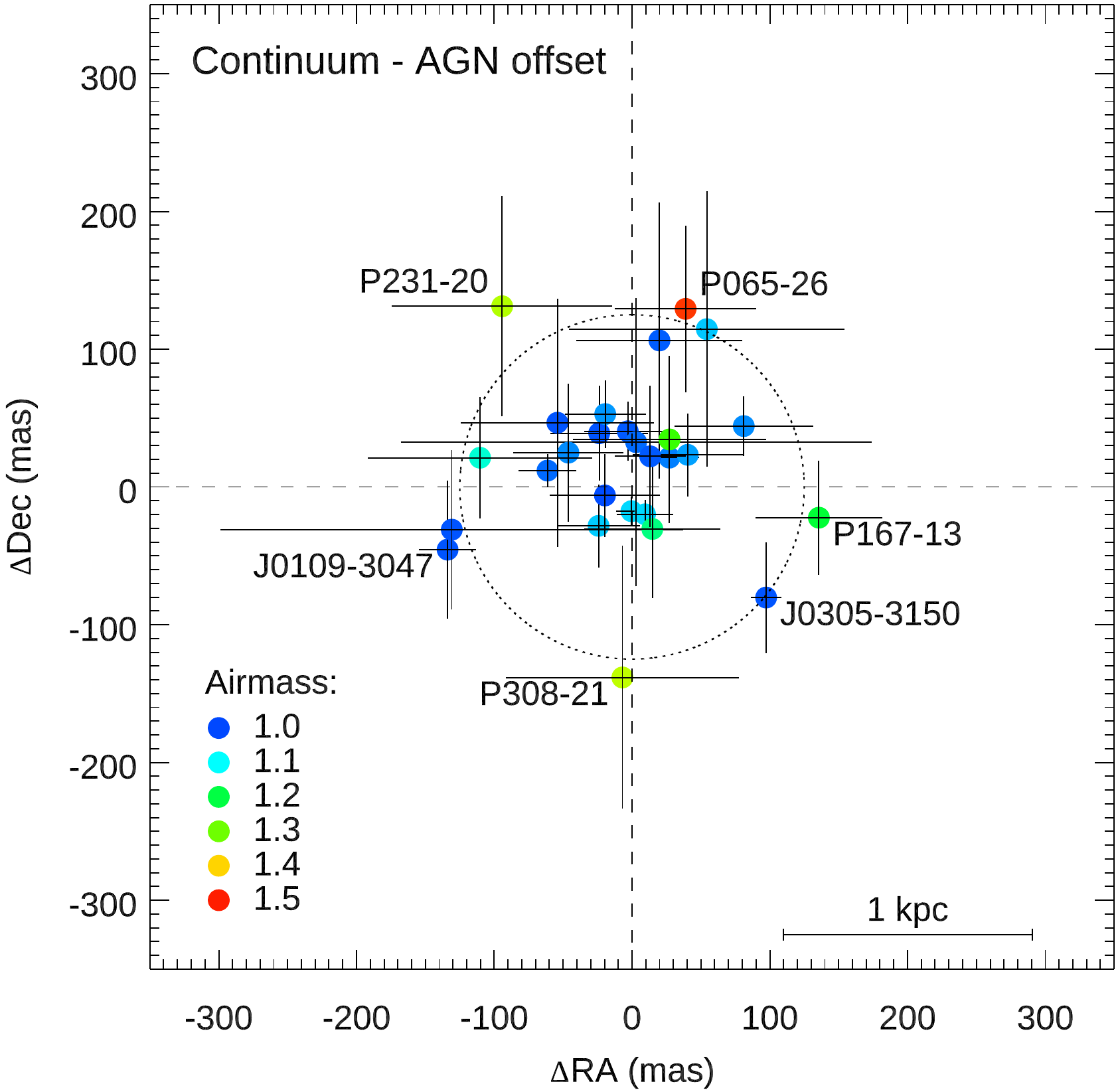} & 
\includegraphics[width=\columnwidth]{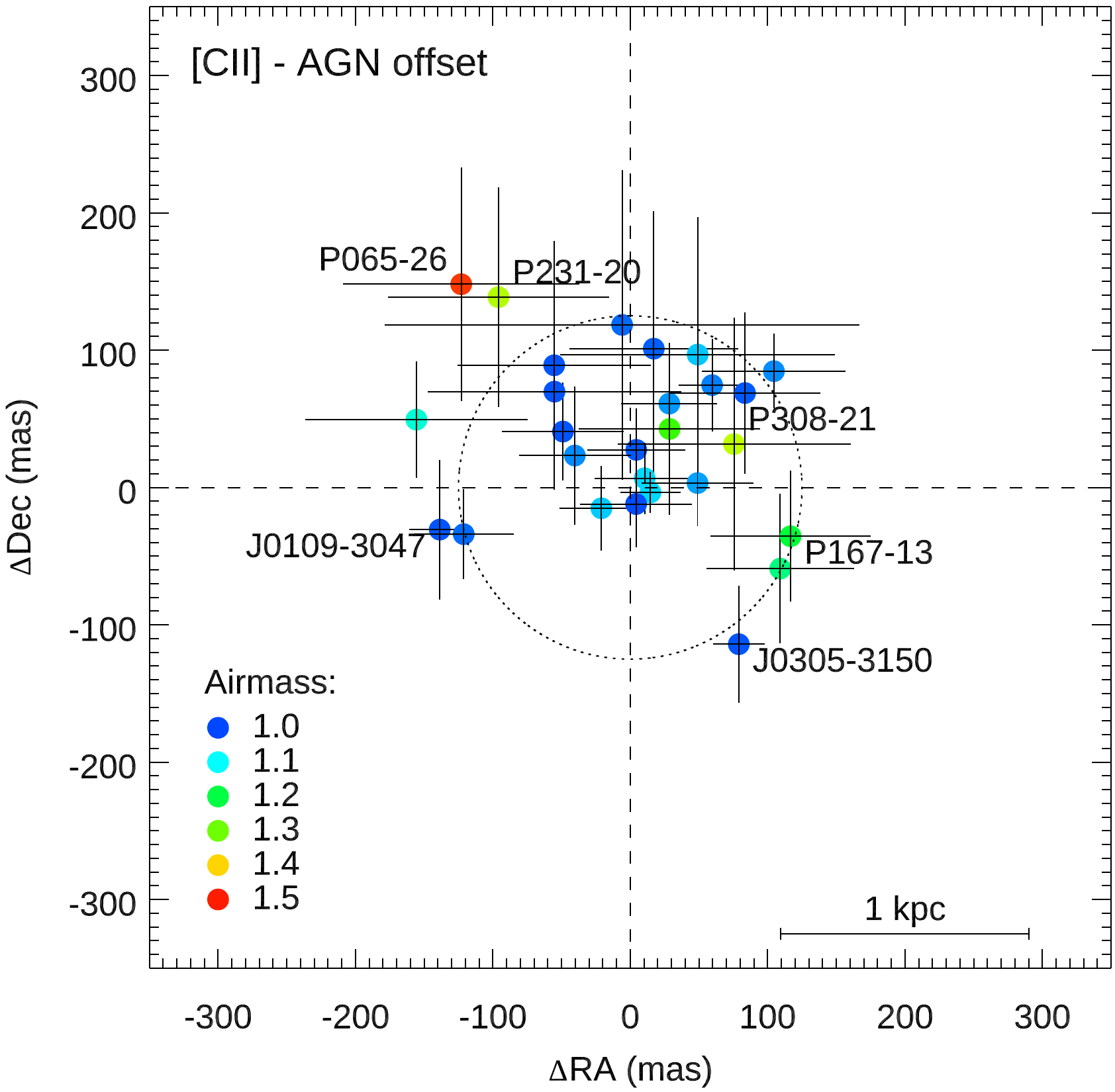}
\end{tabular}
\caption{Left: difference in position on the sky between the accreting black hole (the quasar position) and the center of the 2D Gaussian fit to the continuum emission as reported by the CASA task {\sc imfit}. The dotted circle represents the size of a typical, 0\farcs25 beam. The colors of the points indicate the minimum air mass of the optical/near-infrared imaging of the quasar. Objects with a significant ($>$2\,$\sigma$) offset are labeled. The known merger P308--21 is labeled as well. Right: same as the plot on the left, but here with the central position of the \cii\ emission instead of the dust continuum. \label{fig:position}}
\end{center}
\end{figure*}

\subsection{Line Emitters in the Field}
\label{sec:lineemitters}

\begin{deluxetable*}{lcccccccc}
\tablecaption{Properties of candidate line emitters found in the quasar fields.}
\label{tab:linecandidates}
\tablewidth{0pt}
\tablehead{ \colhead{Name} & \colhead{R.A.\ (ICRS)} &
\colhead{Decl.\ (ICRS)} &
\colhead{$\nu_\mathrm{center}$} &
\colhead{$F$} & \colhead{FWHM} &
\colhead{$S_\mathrm{cont}$} &
\colhead{S/N} & \colhead{Fidelity} \\
\colhead{} & \colhead{} & \colhead{} & \colhead{GHz} &
\colhead{Jy\,\kms} & \colhead{\kms} & \colhead{mJy} &
\colhead{} & \colhead{}}
\startdata
J0100+2802C1 & 01$^\mathrm{h}$00$^\mathrm{m}$14$.\!\!^\mathrm{s}$04 & +28$^\circ$02$^\prime$17$\farcs$4 & 259.49$\pm$0.03 & 0.80$\pm$0.13 & 435$\pm$78 & $<$0.19 & 5.9 & 0.91 \\
P036+03C1 & 02$^\mathrm{h}$26$^\mathrm{m}$00$.\!\!^\mathrm{s}$80 & +03$^\circ$03$^\prime$02$\farcs$4 & 254.78$\pm$0.01 & 0.37$\pm$0.07 & 103$\pm$23 & $<$0.19 & 5.9 & 1.00 \\
P036+03C2 & 02$^\mathrm{h}$26$^\mathrm{m}$01$.\!\!^\mathrm{s}$74 & +03$^\circ$02$^\prime$53$\farcs$5 & 268.79$\pm$0.01 & 0.04$\pm$0.01 & 114$\pm$28 & $<$0.03 & 5.5 & 0.85 \\
J0305--3150C1\tablenotemark{a} & 03$^\mathrm{h}$05$^\mathrm{m}$16$.\!\!^\mathrm{s}$39 & --31$^\circ$50$^\prime$55$\farcs$0 & 249.82$\pm$0.04 & 1.14$\pm$0.29 & 489$\pm$132 & 0.66$\pm$0.13 & 6.0 & 0.96 \\
J0305--3150C2\tablenotemark{a} & 03$^\mathrm{h}$05$^\mathrm{m}$16$.\!\!^\mathrm{s}$87 & --31$^\circ$50$^\prime$55$\farcs$3 & 249.73$\pm$0.01 & 0.21$\pm$0.03 & 121$\pm$22 & 0.16$\pm$0.03 & 7.2 & 1.00 \\
J0305--3150C3\tablenotemark{a} & 03$^\mathrm{h}$05$^\mathrm{m}$16$.\!\!^\mathrm{s}$96 & --31$^\circ$50$^\prime$55$\farcs$8 & 249.42$\pm$0.03 & 0.66$\pm$0.09 & 559$\pm$80 & 0.54$\pm$0.04 & 9.3 & 1.00 \\
J0305--3150C4 & 03$^\mathrm{h}$05$^\mathrm{m}$17$.\!\!^\mathrm{s}$19 & --31$^\circ$50$^\prime$50$\farcs$8 & 252.30$\pm$0.04 & 0.13$\pm$0.06 & 207$\pm$103 & 0.46$\pm$0.04 & 5.9 & 0.94 \\
J0842+1218C1\tablenotemark{b} & 08$^\mathrm{h}$42$^\mathrm{m}$28$.\!\!^\mathrm{s}$97 & +12$^\circ$18$^\prime$55$\farcs$0 & 269.00$\pm$0.01 & 1.81$\pm$0.13 & 331$\pm$26 & $<$0.22 & 18.9 & 1.00 \\
J0842+1218C2\tablenotemark{c} & 08$^\mathrm{h}$42$^\mathrm{m}$29$.\!\!^\mathrm{s}$67 & +12$^\circ$18$^\prime$46$\farcs$3 & 269.04$\pm$0.02 & 0.43$\pm$0.09 & 268$\pm$62 & $<$0.17 & 7.5 & 1.00 \\
J1044--0125C1 & 10$^\mathrm{h}$44$^\mathrm{m}$32$.\!\!^\mathrm{s}$86 & --01$^\circ$24$^\prime$51$\farcs$9 & 294.19$\pm$0.01 & 0.30$\pm$0.06 & 143$\pm$34 & $<$0.17 & 5.7 & 1.00 \\
P167--13C1\tablenotemark{d} & 11$^\mathrm{h}$10$^\mathrm{m}$34$.\!\!^\mathrm{s}$04 & --13$^\circ$29$^\prime$46$\farcs$3 & 253.02$\pm$0.01 & 1.26$\pm$0.08 & 445$\pm$30 & 0.17$\pm$0.04 & 18.8 & 1.00 \\
P183+05C1 & 12$^\mathrm{h}$12$^\mathrm{m}$26$.\!\!^\mathrm{s}$32 & +05$^\circ$05$^\prime$29$\farcs$6 & 255.61$\pm$0.04 & 0.41$\pm$0.12 & 373$\pm$121 & 0.37$\pm$0.06 & 5.7 & 0.92 \\
P183+05C2 & 12$^\mathrm{h}$12$^\mathrm{m}$28$.\!\!^\mathrm{s}$07 & +05$^\circ$05$^\prime$34$\farcs$6 & 240.75$\pm$0.01 & 0.27$\pm$0.06 & 155$\pm$35 & $<$0.13 & 5.8 & 1.00 \\
J1306+0356C1\tablenotemark{c} & 13$^\mathrm{h}$06$^\mathrm{m}$08$.\!\!^\mathrm{s}$33 & +03$^\circ$56$^\prime$26$\farcs$2 & 270.19$\pm$0.01 & 1.29$\pm$0.15 & 200$\pm$26 & 0.35$\pm$0.12 & 11.3 & 1.00 \\
J1319+0950C1 & 13$^\mathrm{h}$19$^\mathrm{m}$11$.\!\!^\mathrm{s}$65 & +09$^\circ$50$^\prime$38$\farcs$2 & 269.50$\pm$0.05 & 0.91$\pm$0.22 & 490$\pm$127 & $<$0.28 & 5.9 & 0.95 \\
J1319+0950C2 & 13$^\mathrm{h}$19$^\mathrm{m}$11$.\!\!^\mathrm{s}$87 & +09$^\circ$50$^\prime$44$\farcs$3 & 266.02$\pm$0.02 & 0.35$\pm$0.08 & 218$\pm$55 & 0.19$\pm$0.05 & 5.6 & 0.90 \\
J1342+0928C1 & 13$^\mathrm{h}$42$^\mathrm{m}$08$.\!\!^\mathrm{s}$24 & +09$^\circ$28$^\prime$43$\farcs$4 & 222.70$\pm$0.02 & 0.10$\pm$0.03 & 222$\pm$75 & $<$0.06 & 5.8 & 1.00 \\
J1342+0928C2 & 13$^\mathrm{h}$42$^\mathrm{m}$08$.\!\!^\mathrm{s}$65 & +09$^\circ$28$^\prime$44$\farcs$2 & 238.84$\pm$0.04 & 0.69$\pm$0.16 & 475$\pm$117 & 1.76$\pm$0.07 & 24.3 & 1.00 \\
P231--20C1 & 15$^\mathrm{h}$26$^\mathrm{m}$37$.\!\!^\mathrm{s}$11 & --20$^\circ$49$^\prime$59$\farcs$4 & 251.86$\pm$0.01 & 0.19$\pm$0.04 & 145$\pm$38 & $<$0.11 & 6.0 & 0.96 \\
P231--20C2 & 15$^\mathrm{h}$26$^\mathrm{m}$37$.\!\!^\mathrm{s}$62 & --20$^\circ$49$^\prime$58$\farcs$6 & 235.04$\pm$0.01 & 0.12$\pm$0.02 & 209$\pm$45 & $<$0.05 & 6.2 & 1.00 \\
P231--20C3\tablenotemark{b} & 15$^\mathrm{h}$26$^\mathrm{m}$37$.\!\!^\mathrm{s}$87 & --20$^\circ$50$^\prime$02$\farcs$3 & 250.38$\pm$0.01 & 2.81$\pm$0.22 & 497$\pm$41 & 1.52$\pm$0.10 & 34.0 & 1.00 \\
P231--20C4\tablenotemark{c} & 15$^\mathrm{h}$26$^\mathrm{m}$37$.\!\!^\mathrm{s}$97 & --20$^\circ$50$^\prime$02$\farcs$5 & 250.38$\pm$0.03 & 0.25$\pm$0.05 & 334$\pm$77 & 0.10$\pm$0.03 & 6.1 & 1.00 \\
J2054--0005C1 & 20$^\mathrm{h}$54$^\mathrm{m}$07$.\!\!^\mathrm{s}$01 & --00$^\circ$05$^\prime$25$\farcs$7 & 270.03$\pm$0.01 & 0.55$\pm$0.15 & 106$\pm$32 & $<$0.46 & 5.8 & 0.92 \\
J2100--1715C1\tablenotemark{b} & 21$^\mathrm{h}$00$^\mathrm{m}$55$.\!\!^\mathrm{s}$45 & --17$^\circ$15$^\prime$22$\farcs$1 & 268.42$\pm$0.03 & 8.09$\pm$1.07 & 600$\pm$80 & 3.66$\pm$0.46 & 15.4 & 1.00 \\
J2318--3113C1 & 23$^\mathrm{h}$18$^\mathrm{m}$18$.\!\!^\mathrm{s}$72 & --31$^\circ$13$^\prime$49$\farcs$9 & 256.50$\pm$0.01 & 0.09$\pm$0.02 & 102$\pm$26 & $<$0.06 & 5.7 & 1.00 \\
J2318--3113C2 & 23$^\mathrm{h}$18$^\mathrm{m}$19$.\!\!^\mathrm{s}$36 & --31$^\circ$13$^\prime$48$\farcs$9 & 254.85$\pm$0.01 & 0.17$\pm$0.05 & 111$\pm$35 & $<$0.13 & 5.6 & 0.99 \\
J2318--3029C1 & 23$^\mathrm{h}$18$^\mathrm{m}$33$.\!\!^\mathrm{s}$36 & --30$^\circ$29$^\prime$44$\farcs$5 & 266.84$\pm$0.04 & 0.93$\pm$0.21 & 427$\pm$104 & $<$0.33 & 6.0 & 0.91 \\
\enddata
\tablenotetext{a}{Previously known companion, published in \citet{ven19}.}
\tablenotetext{b}{Previously known companion, published in \citet{dec17}.}
\tablenotetext{c}{Previously known companion, published in \citet{nee19}.}
\tablenotetext{d}{Previously known companion, published in \citet{wil17}.}
\end{deluxetable*}

To identify potential new companion galaxies, we searched a total of 56 original (i.e., not continuum-subtracted) data cubes for galaxies with an emission line. The number of data cubes was two to three per quasar field, one covering the two bandpasses around the redshifted \cii\ emission line of the quasar and one or two covering the continuum bandpasses (see Section~\ref{sec:obs}). We followed a procedure that is similar to that described in \citet{gon19} but with a few differences.
We searched the cubes by channel for pixels reaching an S/N threshold of 3. To avoid including the same source multiple times, we remove candidates within 0\farcs5 and 1000\,\kms\ of the highest-S/N candidates. To search for broader lines, we convolved the data cubes along the frequency axis with a Gaussian kernel with a standard deviation increasing incrementally by 30 MHz up to 400 MHz (approximately 500\,\kms). We combined the results of each convolution and again removed sources within the distance and frequency thresholds (0\farcs5 and 1000\,\kms, respectively) to avoid double counting the same line candidates. 

To assign fidelity estimates to our list of candidates, we first repeated the line search process except we inverted the cube, searching for the brightest negative peaks at various line widths. For each line width, we binned the detected positive and negative candidates by S/N ratio and computed the fidelity:

\begin{equation}
\label{eqn:fidelity}
    f = 1 - \frac{n_{\text{negative}, i}}{n_{\text{positive}}, i},
\end{equation}

\noindent
where $n_{\text{negative}, i}$ and $n_{\text{positive}, i}$ are the numbers of negative and positive candidates in an S/N bin $i$. For our search, we only considered candidates with a fidelity $\geq 0.8$. 

The line search code yielded a total of 403 (305) candidate line emitters with a fidelity above 0.8 (0.9) in 56 data cubes. We visually inspected all candidates and discarded the ones that were dominated by continuum emission or extended line emission from the quasar (136 sources) or by imaging artifacts such as side lobes (54 sources). We fitted the spectra of the remaining 213 candidates with a Gaussian profile. We rejected 50 candidates with a line width of $\lesssim$50\,\kms\ (i.e.\ a single channel of emission), as these are likely to be noise. We subsequently created channel maps both centered on and away from the line emission (see Appendix~\ref{appendix} for details) and removed 136 objects for which the imaging showed issues, such as the emission falling on top of severe striping. This left 27 candidate line emitters (of which 26 had fidelity $>$0.9). 

We confirmed 10 previously detected emission line sources \citep{dec17,wil17,nee19,ven19} and present 17 new candidate line emitters. The properties of all emission line candidates are listed in Table~\ref{tab:linecandidates}. We further discuss the candidates that are likely associated with the quasars in Section~\ref{sec:clustering}.

\section{Discussion}
\label{sec:discussion}

\subsection{Quasar Location}
\label{sec:location}

\begin{figure}
\begin{center}
\includegraphics[width=\columnwidth]{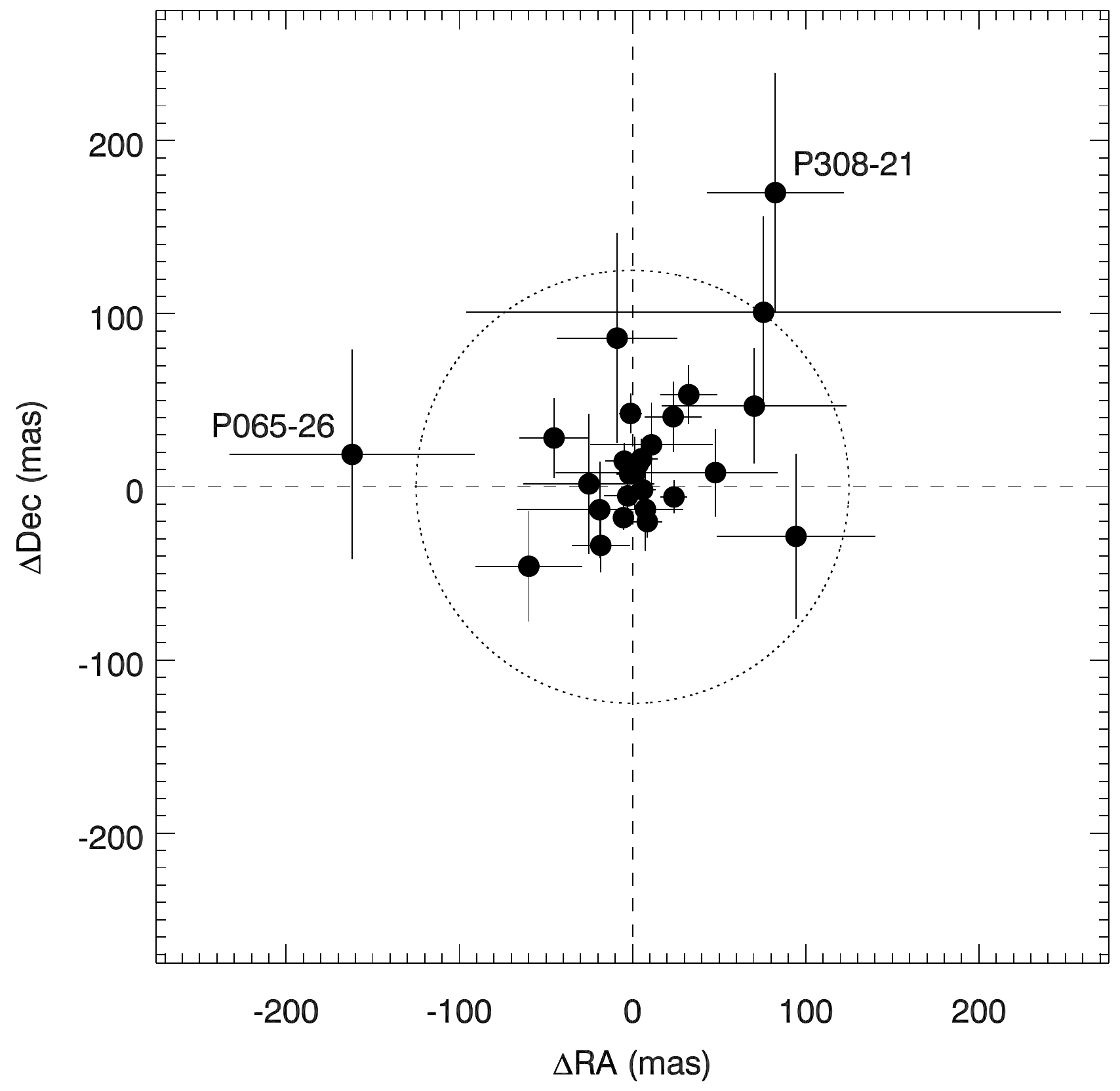}
\caption{Offset between the center of the dust and that of the \cii\ emission. For most of the quasar host galaxies in our sample, there are no significant offsets.}
\label{fig:ciicontoffset}
\end{center}
\end{figure}

It is typically assumed that the supermassive black hole of a quasar host galaxy is located at the very center of the gravitational potential of a given dark matter halo. The gaseous component (that partly fuels the supermassive black hole) is also located toward this gravitational center. This is also seen in simulations, where the central regions of dark matter halos are those where the infalling gas accumulates. The exact position of the black hole is typically put in `by hand' at the bottom of the potential well in most simulations \citep[e.g.,][]{wei17}, although not all \citep[e.g.,][]{bar20}. It is therefore interesting to see if the gas (and dust) seen in the quasar hosts are indeed cospatial and centered on the supermassive black hole. With accurate quasar positions now available through matching to the Gaia DR2 catalog, leading to positional accuracies of $<$0\farcs15 (Section~\ref{sec:astrometry}), and $<$0\farcs3 imaging of the \cii\ line and dust continuum, we now have the measurements in hand to characterize any possible offsets between the gas and the central black hole. Any offset in position could be due to a number of things, such as infalling gas that has not fully settled in a disk, the presence of mergers that shift the center of mass in the quasar host, or the possibility that the supermassive black hole is in fact {\em not} located at the precise center of the potential well.

The locations of the accreting black holes are listed in Table~\ref{tab:sample} and the central positions and uncertainty of the dust continuum and \cii\ emission are the centroid position obtained by {\sc imfit}. The results are shown in Figure~\ref{fig:position}. For most sources, the center of the dust emission coincides with the optical/near-infrared point source to within 0\farcs1. Although in most cases not significant, the predominantly positive offset in decl.\ could be due to differential chromatic refraction when calculating the optical positions (see Section~\ref{sec:astrometry}), but the effect seems to be much smaller than the typical beam of 0\farcs25 of our ALMA observations. As gas traced by the \cii\ emission extends to large radii in the quasar hosts, we also compared the position of the quasars with the \cii\ position (the centroid provided by {\sc imfit}) in Figure~\ref{fig:position}. Although the offsets are on average slightly larger, still the majority of sources do not show a significant offset. Several sources with known nearby companions, P231--20, P167--13, J0305--3150, and P308--21, show offsets between the quasar and the dust emission in the host galaxy. Intriguingly, two sources without known companions, P065--26 and J0109--3047, show significant offsets between the optical position of the central quasar and that of both the dust and \cii\ emission. For J0109--3047, a large velocity offset of $\sim$1700\,\kms\ between the \cii\ redshift and that of the broad emission lines near the black hole was reported \citep[e.g.,][]{ven16}, which, combined with the spatial offsets presented here, could indicate that the quasar host galaxy has recently undergone an interaction with a different system. Alternatively, these offsets could be due to an outflow. Higher-spatial-resolution observations are required to distinguish between these two scenarios. The offsets measured for P065--26 could be, in part, due to systemic astrometric uncertainties as this source has been observed at high air mass. However, when comparing the position of the dust and the \cii\ emission (Figure~\ref{fig:ciicontoffset}), P065--26 has the largest offset, together with the known merging system P308--21. Also, the \cii\ emission of P065--26 appears to have an irregular, disturbed morphology (Figure~\ref{fig:allmaps}).

We conclude that there is overall excellent agreement between the location of the supermassive black hole and the ISM of the quasar host, seen in both \cii\ and dust continuum emission. In most cases, the offset is less than a few hundred parsecs, but some outliers with offsets of $\sim$750\,pc exist. Those outliers can in many cases be linked to host galaxies that undergo interactions, which may lead to a displacement of the ISM due to gravitational interactions. For some sources, we cannot discard that the offsets are caused by outflows. The presence of outflows in our sample has been extensively discussed in \citet{nov20}. Overall we conclude that, to first order, the center of mass of the ISM is cospatial (within the observational uncertainties of a few 100\,pc) with the central supermassive black hole.

\subsection{Star Formation Rate Density}
\label{sec:sfrd}

\begin{figure}
\begin{center}
\includegraphics[width=\columnwidth]{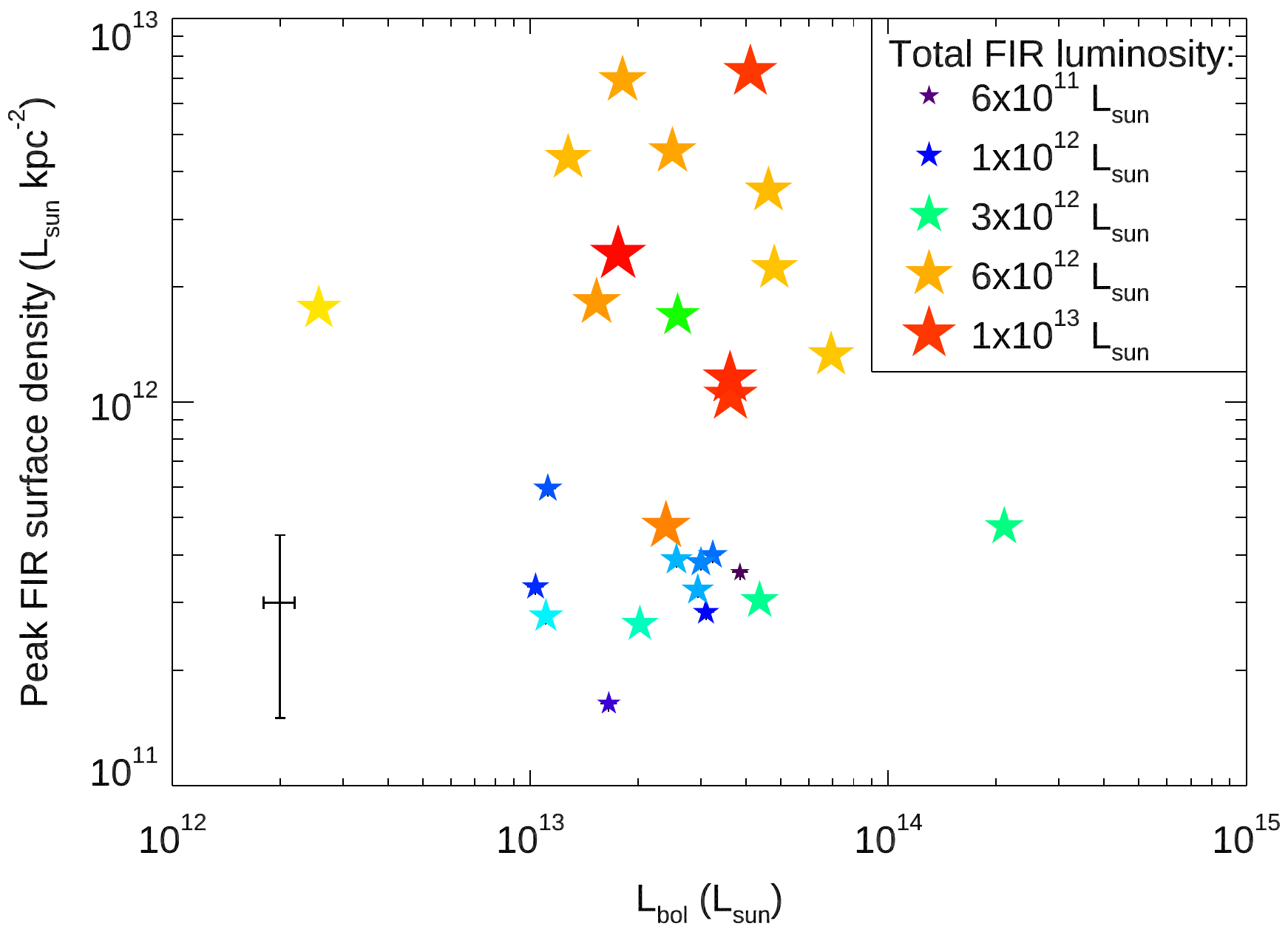}
\caption{Peak FIR surface density vs.\ bolometric luminosity. The peak FIR luminosities were derived from the peak pixel in the continuum maps (Figure~\ref{fig:allmaps}). The color of each point represents the {\em total} FIR luminosity of the quasar host as listed in Table~\ref{tab:results}. A typical error bar is shown in the lower left corner. There is no correlation between the brightness of the accreting black hole and the dust continuum peak flux density.}
\label{fig:peakbol}
\end{center}
\end{figure}

An important question arising in FIR studies of high-redshift quasar host galaxies is the heating source of the dust. Models suggest that the high luminosity of the central quasar could potentially power a significant fraction of the dust emission for even the most FIR-luminous quasars \citep[e.g.,][]{sch15}. Recently, \citet{ven18} compared the quasar's bolometric luminosity and FIR luminosity of a large sample of $z>5.7$ quasars and found only a weak correlation. However, the FIR luminosities used in \citet{ven18} were mostly derived from unresolved observations. Therefore, it remains possible that the quasar still contributes significantly to the dust heating close to the accreting black hole. With the kiloparsec-resolution observations presented in this paper, we test whether this is the case for our sample of quasars. We measured the peak flux density in our continuum maps (Figure~\ref{fig:allmaps}) and converted these values to FIR luminosities using Eq.~\ref{eq:fir}. The peak FIR surface densities are between $10^{11}$ and $10^{13}$\,\lsun\,kpc$^{-2}$. The bolometric luminosity $L_\mathrm{bol}$ of the quasars was taken from the literature \citep[Table 3 in][]{ven18}. In short, the quasar's bolometric luminosity is computed based on the rest-frame UV emission of the accreting black hole. We note that the quasars studied here are unobscured (Type I) quasars and show no signs of dust absorption in their UV spectra. 
We compared $L_\mathrm{bol}$ to the peak FIR surface density in Figure~\ref{fig:peakbol}. The brightness of the central dust emission seems to be independent of the luminosity of the central active galactic nucleus (the Pearson correlation coefficient is $r=-0.04$). This suggests that even in the central kiloparsec of the hosts of luminous $z\sim6$ quasars, the dust is predominantly heated by star formation. This is somewhat surprising, given that the accreting black hole, with a bolometric luminosity often a factor of 10 higher than the FIR luminosity, is sitting at the center of a dusty galaxy. The lack of heating by the AGN could be explained if the emission from the quasar is highly collimated and emitted perpendicular to the disk of the host galaxy.  Alternatively, the obscuration of the central black hole could be anisotropic. With a large fraction of quasar host galaxies having a very close companion or undergoing a merger event (Section~\ref{sec:maps}), the distribution of dust in the centers of the quasar hosts could show significant variation between sources, which in turn could erase any correlation between the emission of the accreting black hole and that of the surrounding dust.

\begin{figure}
\begin{center}
\includegraphics[width=\columnwidth]{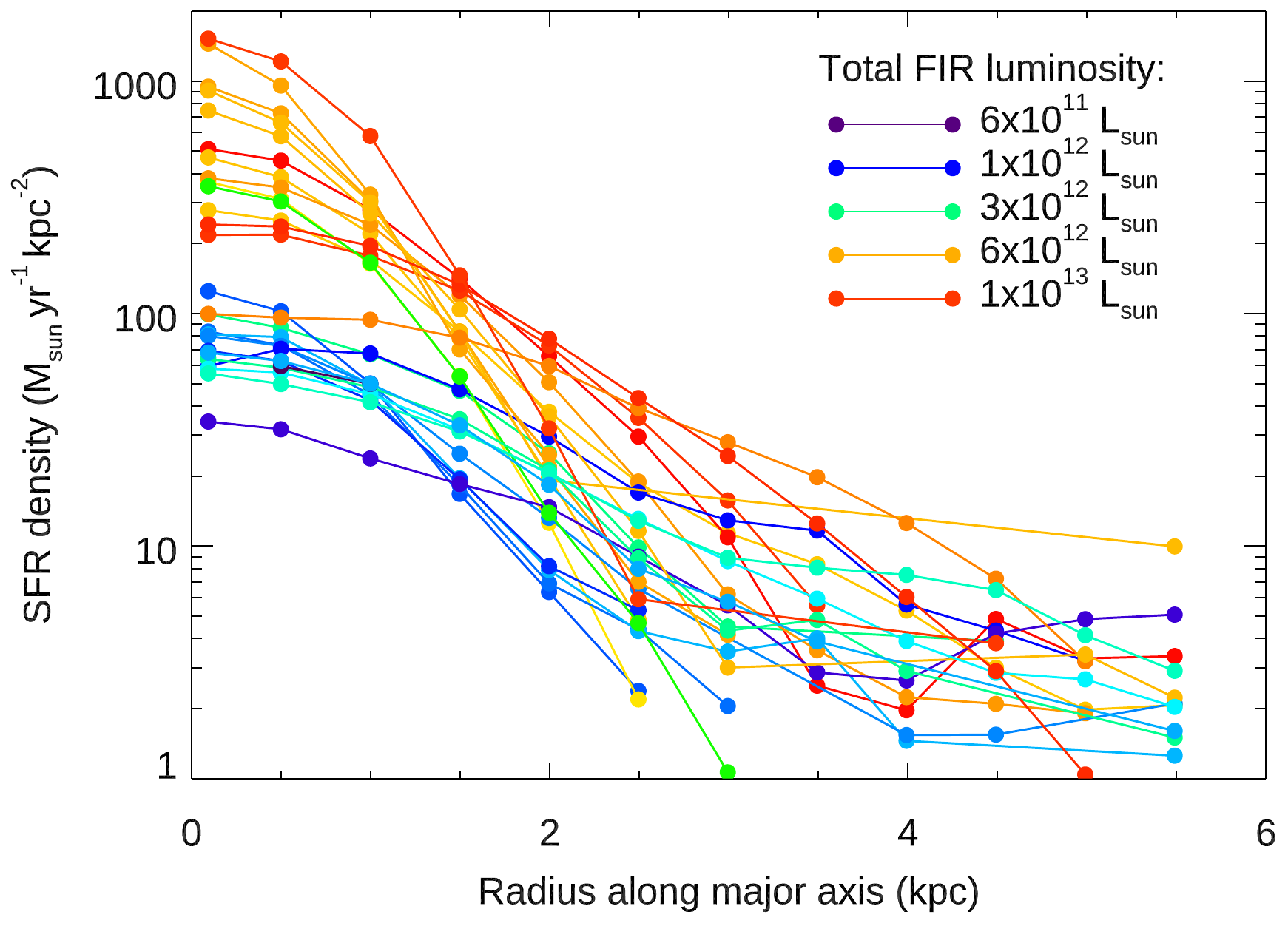}
\caption{SFR surface density in the quasar host galaxies as a function of radius from the center. The SFR was derived from the dust continuum flux density (see text for details). Each quasar host galaxy is represented by a colored track with the color representing the total FIR luminosity of the source (Table~\ref{tab:results}). Most peak star formation rate densities are below the Eddington limit for star formation (i.e.\ $\lesssim$\,1000\,\msunyr\,kpc$^{-2}$).}
\label{fig:sfrd}
\end{center}
\end{figure}

In Figure~\ref{fig:sfrd} we show the SFR surface density (SFRD) in annuli around the center of the quasar host galaxy (defined as the peak of the dust continuum emission), derived from Eqs.~\ref{eq:fir} and \ref{eq:sfr}. The peak SFRD spans a large range of nearly a factor of 50, from 30\,\msunyr\,kpc$^{-2}$ to 1500\,\msunyr\,kpc$^{-2}$. This range is larger than that of the total FIR luminosities covered by our sample (spanning a factor of $\sim$25; see Table~\ref{tab:results}). While the highest SFRDs that we find approach that of the Eddington limit for star formation \citep[e.g.,][]{wal09b}, as also seen in the centers of local ultraluminous infrared galaxies \citep[ULIRGs; e.g.,][]{dow98} or Galactic star forming regions such as Orion, most quasar host galaxies have central SFRD values well below such an  extreme value (i.e., $\lesssim$1000\,\msunyr\,kpc$^{-2}$). In the outskirts, SFRD values reach a few \msunyr\,kpc$^{-2}$, which is still significantly higher than typical star formation rate densities in nearby disk galaxies \cite[i.e., $<\,1$\msunyr\,kpc$^{-2}$,][]{ler13}.

\subsection{The \cii-to-FIR Ratio and \cii\ Deficit}
\label{sec:ciifir}

\begin{figure}
\begin{center}
\includegraphics[width=\columnwidth]{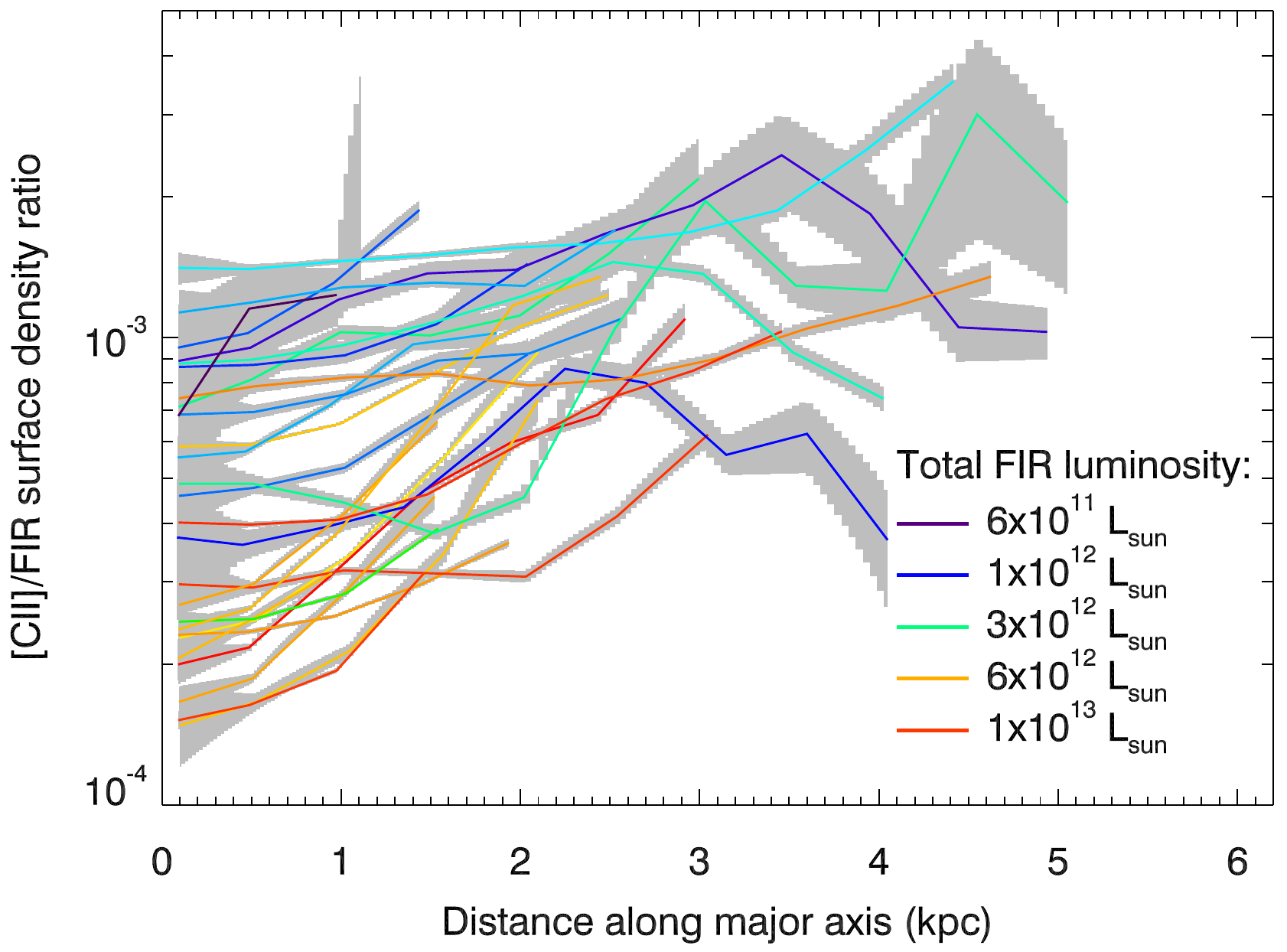}
\caption{\cii-to-FIR surface density luminosity ratio of our quasar sample as a function of distance from the center. The colors encode the total FIR luminosity of the sources (Table~\ref{tab:results}), and the gray regions represent the 1\,$\sigma$ uncertainties in the luminosity ratio. On average, the \cii/FIR ratio is lowest in the center of the galaxies.}
\label{fig:ciifirrad}
\end{center}
\end{figure}

Studies of local ULIRGs show that their centers often display a low \cii/FIR luminosity ratio, which is attributed to intense radiation fields in the nucleus \citep[e.g.,][]{smi17,her18}. In Figure~\ref{fig:ciifirrad} we show that generally the \cii-to-FIR luminosity ratio is lower in the center of the quasar hosts, although the central values vary up to a factor of $\sim$10 within our sample. In the centers, we find surface density values as low as a few $\times 10^{-4}$, consistent with studies of $z>1$ starburst galaxies \citep[e.g.,][]{lam18,lit19,ryb20}. Away from the central regions, at radii $>$2\,kpc, the \lcii/\lfir\ ratio approaches luminosity ratios of $\sim$$10^{-3}$, similar to what is found in the disks of local starburst galaxies \citep[e.g.,][]{dia13,sar14}. 

\begin{figure}[t]
\begin{center}
\includegraphics[width=\columnwidth]{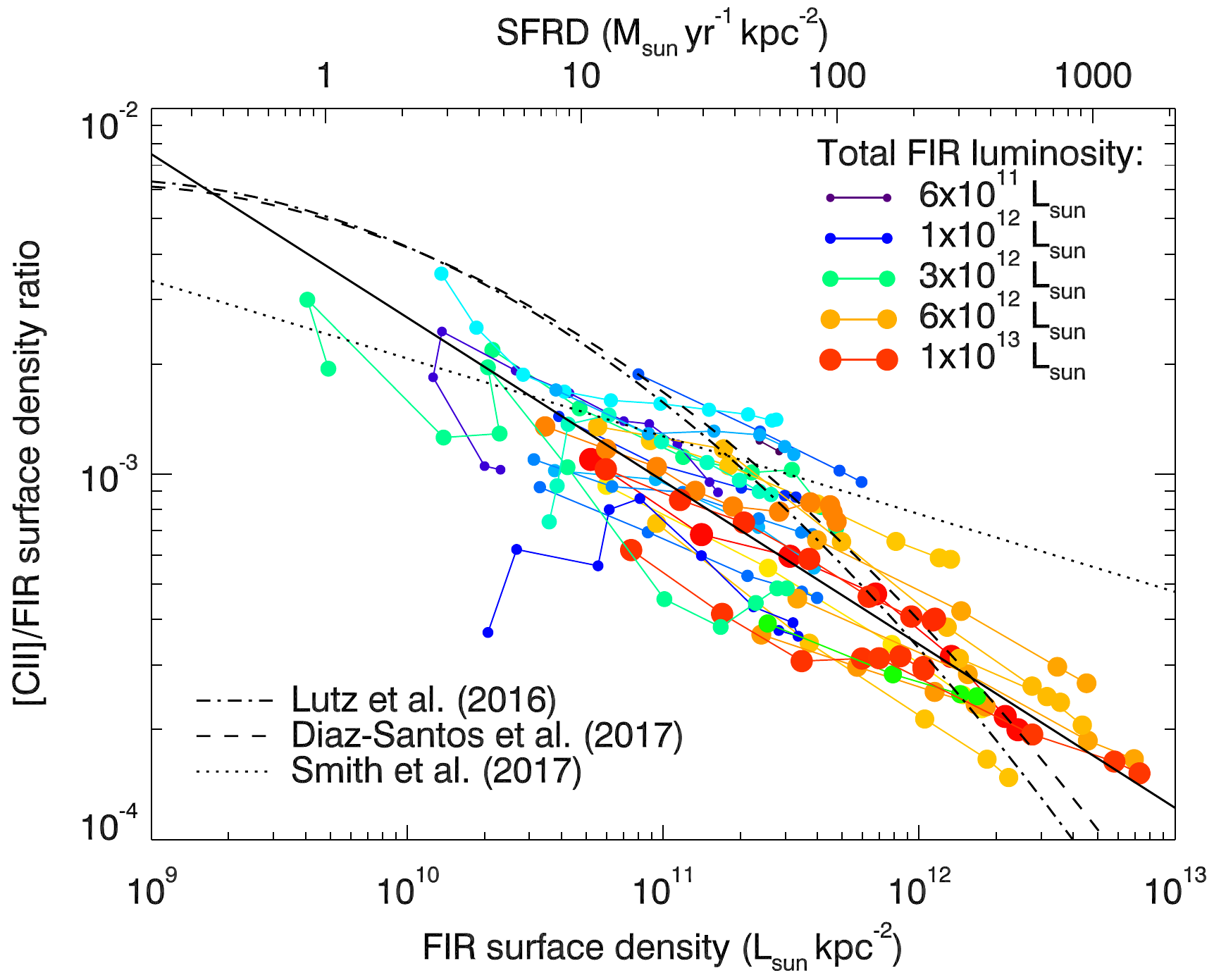}
\caption{\cii/FIR surface density luminosity ratio, now shown as a function of FIR luminosity density. The colors of the tracks again indicate the total FIR luminosity of the objects. Overplotted are relations between the \cii/FIR surface densities and the FIR surface density from the literature, after converting their measurements to the assumptions and units used in this work (see legend for details). Our best-fit power-law fit to the data points is shown as a solid line.}
\label{fig:ciifirfir}
\end{center}
\end{figure}

To investigate the importance of the FIR surface density (i.e.\ the local radiation field) in driving the low \cii/FIR luminosity ratios, we plot the \lcii/\lfir\ surface densities as a function of the FIR surface density $\Sigma_\mathrm{FIR}$ in Figure~\ref{fig:ciifirfir}. 
We find that the lowest \lcii/\lfir\ ratios come from the regions that show the highest \lfir\ values, that is, in the centers. This indicates that the intense radiation field in the centers of the quasar host galaxies (not necessarily due to the radiation of the accreting black hole; see Section~\ref{sec:sfrd}) could be the cause of the \cii\ deficit, similar to the results found for local starburst galaxies and quasars \citep[e.g.,][]{lut16,smi17,her18}. 

We fitted a power-law to the data in Figure~\ref{fig:ciifirfir} and found the following relation:

\begin{equation}
\Sigma_\mathrm{[CII]/FIR} = 9.6\times10^{-4} \left(\frac{\Sigma_\mathrm{FIR}}{10^{11} L_\odot\,\mathrm{kpc}^{-2}}\right)^{-0.447}
\end{equation}

The slope of this correlation is very similar to the one published by \citet{dia14}, who found a slope of $-0.39\pm0.07$ for the extended regions in spatially resolved local (ultra)luminous infrared galaxies. This slope is significantly shallower than predicted by models, possibly caused in part by the reduced gas heating efficiency due to the ionization of dust particles \citep[see][for an extensive discussion]{dia14}.
Our data in Figure~\ref{fig:ciifirfir} are below the relation presented by \citet{smi17}. However, the data fitted by these authors were regions inside nearby galaxies, where FIR surface densities are a few orders of magnitude lower than those studied here. Both \citet{lut16} and \citet{dia17} also presented \lcii/\lfir\ ratios as a function of FIR surface density. Generally, they find a steeper relation with higher ratios at low FIR surface densities compared to our $z\sim6$ quasar host galaxy sample, whereas at the highest densities, their fits predict lower values than reported here. These discrepancies could be due to our assumption of a single dust temperature everywhere in the quasar host galaxies, whereas studies of local galaxies \citep[e.g.,][]{dia17} report a range of dust temperatures. Furthermore, it has been shown in local galaxies that there is a relation between the infrared surface brightness and the dust temperature \citep[e.g.,][]{cha07}.
We note that the estimation of $\Sigma_\mathrm{FIR}$ is sensitive to the temperature that is used in the derivation of the luminosity. If the outskirts of our quasar hosts have lower dust temperatures compared to the centers, then our \cii/FIR luminosity ratio would increase at low FIR surface densities. In particular, if the outskirts of the quasar host galaxies where $\Sigma_\mathrm{FIR} \approx 2\times10^{10}$\,\lsun\,kpc$^{-2}$ have a lower dust temperature of $T_d = 35$\,K \citep[well within the range of dust temperatures spanned by local starburst galaxies; e.g.,][]{dia17}, then $\Sigma_\mathrm{FIR}$ will be a factor of $\sim$2 lower, while $\Sigma_\mathrm{[CII]/FIR}$ would be higher by this factor. This would place the outskirt of our quasar hosts on the local relations as found by \citet{lut16} and \citet{dia17}. Similarly, if the centers of the quasar hosts are hotter, the \cii-to-FIR ratio would decrease for the highest FIR surface densities. Future observations of the quasar hosts with ALMA band~8 or band~9 will shed light on this question.

\subsection{Galaxy Clustering around High-redshift Quasars}
\label{sec:clustering}

\begin{deluxetable*}{lccccc}
\tablecaption{Newly Discovered Companion Galaxies in the Quasar Fields.}
\label{tab:companions}
\tablewidth{0pt}
\tablehead{\colhead{Name} & \colhead{$z_\mathrm{[CII]}$} &
\colhead{\lcii} & \colhead{\lfir} &
\colhead{Distance} & \colhead{$\Delta v$} \\
\colhead{} & \colhead{} &
\colhead{($10^8$\,\lsun)} & \colhead{($10^{11}$\,\lsun)} &
\colhead{(kpc)} & \colhead{(\kms)}}
\startdata
J0100+2802C1 & 6.3241$\pm$0.0008 & 8.12$\pm$1.36 & $<$4.05 & 87.8 & --110$\pm$30 \\
P183+05C1 & 6.4352$\pm$0.0012 & 4.26$\pm$1.28 & 8.13$\pm$1.40 & 57.3 & --140$\pm$50 \\
J1319+0950C2 & 6.1444$\pm$0.0005 & 3.41$\pm$0.77 & 3.78$\pm$1.00 & 63.2 & +410$\pm$20 \\
J1342+0928C1 & 7.5341$\pm$0.0009 & 1.29$\pm$0.39 & 1.69$\pm$0.56 & 27.2 & --210$\pm$30 \\
P231--20C1 & 6.5459$\pm$0.0004 & 2.03$\pm$0.48 & $<$2.51 & 56.6 & --1620$\pm$20 \\
J2054--0005C1 & 6.0383$\pm$0.0003 & 5.28$\pm$1.41 & $<$9.05 & 77.3 & --20$\pm$10 \\
J2318--3113C1 & 6.4095$\pm$0.0003 & 0.92$\pm$0.21 & $<$1.33 & 31.5 & --1340$\pm$10 \\
J2318--3113C2 & 6.4575$\pm$0.0004 & 1.74$\pm$0.48 & $<$2.97 & 71.0 & +590$\pm$10 \\
J2318--3029C1 & 6.1223$\pm$0.0010 & 9.01$\pm$2.08 & 6.61$\pm$2.20 & 64.7 & --980$\pm$40 \\
\enddata
\end{deluxetable*}

In Section~\ref{sec:lineemitters} we presented 27 candidate line emitters found in the fields of the 27 quasars in our sample. Ten of those were previously identified and published companion \cii-emitting galaxies to the quasars \citep[see][]{dec17,wil17,nee19,ven19} with physical distances up to $\sim$60\,kpc and line-of-sight velocities $\Delta v<600$\,\kms. We note that we did not recover the known companion next to P308--21 \citep{dec17,dec19}, because the quasar host galaxy is actively undergoing a merger and has very extended \cii\ emission (see also Table~\ref{tab:size}).
For the remaining 17 line candidates, we cannot determine a priori if they are \cii\ emitters and whether they are associated with the quasar. For each line, we computed the redshift if the line was \cii\ and calculated the line-of-sight velocity to the quasar. 
In the literature, velocity differences between 1000\,\kms\ and 3000\,\kms\ are typically used to identify sources that are associated with active galactic nuclei \citep[e.g.,][]{hen06,ven06,ven07a,ell10}.
We therefore conservatively assume that sources with a velocity difference of $\Delta v < 2000$\,\kms\ are associated with the quasar. Of the 27 emission line candidates, 19 (including the 10 known companions) are \cii-emitting companion galaxies to the quasars, with 17 of those within 1000\,\kms\ of the quasar. We list the nine new companion galaxies with derived line and continuum properties in Table~\ref{tab:companions}, and we show the spectra and emission maps in Figure~\ref{fig:linecandidatemaps}.

The next question is whether these companion sources represent an overdensity of galaxies around these high-redshift quasars. Spatially, there is no significant difference between the new companion galaxies and the other candidate line emitters, especially if the already-known, nearby ($<$10\,kpc) companions are excluded. The fraction of all channels in our 56 data cubes that probe \cii\ with a velocity difference of $<2000$\,\kms\ with respect to the quasar redshift is 38\%. For $\Delta v < 1000$\,\kms, this fraction is 22\%. Based on the ``field'' density of line emitters (eight emission line candidates that have $\Delta v > 2000$\,\kms), we would only expect $\sim$5 ($\sim$3) emission line candidates with $\Delta v < 2000$\,\kms\ ($\Delta v < 1000$\,\kms), where we find 19 (17). 

This suggests that the emission line candidates are highly clustered around the redshift of the quasar, implying that the majority of sources within 2000\,\kms\ are likely companion galaxies at similar redshifts. Assuming that all candidate line emitters are high-redshift \cii-emitting galaxies, we calculated the average overdensity of quasar companions using 

\begin{equation}
\label{eq:delta}
\delta=(n_\mathrm{obs}-n_\mathrm{exp})/n_\mathrm{exp}, 
\end{equation}

\noindent
where $n_\mathrm{obs}$ and $n_\mathrm{exp}$ are the observed and expected number of galaxies within 2000\,\kms, respectively. We find an overdensity of $\delta = 2.4$ for companions with $\Delta v < 2000$\,\kms. This increases to $\delta \approx 4.3$ for the companions with $\Delta v < 1000$\,\kms. 
We stress that these overdensities are strict lower limits as it is expected that the majority of line candidates with $\Delta v > 2000$\,\kms\ are foreground CO lines and not high-redshift \cii\ emitters \citep[e.g.,][]{dec20}. In this case, it will lower $n_\mathrm{exp}$ in Eq.~\ref{eq:delta} and thus increase $\delta$. Our results therefore confirm and extend the clustering measurement by \cite{dec17} based on a smaller sample of $z\sim6$ quasars with companions.

\section{Summary and Conclusions}
\label{sec:summary}

We present a summary of the \cii\ 158\,$\mu$m emission line and underlying far-infrared (FIR) continuum of 27 quasar host galaxies at $z\sim6$ observed with ALMA at a spatial resolution of $\sim$1\,kpc. The sample covers about one-half of the quasars currently detected with ALMA in the \cii\ emission at these redshifts, with a slight bias to the brighter sources in the complete sample: the mean (median) \cii\ luminosity of our sample is $3\times10^9$ ($2.1\times10^9$)\,\lsun, whereas that of all ALMA-detected quasars is $2.4\times10^9$ ($1.5\times10^9$)\,\lsun. The spatial resolution of our sample allows us to spatially resolve the emission in the centers of the quasar host galaxies. We find that the \cii\ emission in the quasar hosts has sizes between 1.0 and 4.8\,kpc. We note that this holds for the bright centers of the galaxies and does not account for a more extended \cii\ component that can only be quantified by stacking the emission of all quasars in our sample \citep{nov20}. The dust emission is more centrally concentrated than the \cii\ in the central regions. Like in the case of \cii, a more extended faint dust component can only be recovered through stacking \citep{nov20}. 

We find that 13 quasars (48\%) have companion \cii-emitting galaxies in the field (only considering candidates with a velocity difference up to 2000\,\kms), with distances between 3 and 88\,kpc. We visually inspected the maps of the \cii\ emission for nearby $<$5\,kpc companion galaxies or irregular morphologies and
conclude that at least eight quasar hosts are undergoing a merger or have a disturbed \cii\ morphology at the spatial scales probed by our observations. 

Both the interstellar medium of the quasar host galaxies and their accreting supermassive black holes are thought to be located at the bottom of the potential wells of their respective dark matter halos. The high resolution of the ALMA observations, together with improved positions of the central accreting supermassive black holes based on Gaia DR2 corrections, now allow us to observationally test this hypothesis. We find that the interstellar medium and the central black holes are indeed cospatial (with offsets $\lesssim$0\farcs1) in most cases. This suggests that the majority of the central black holes, as well as the interstellar medium, are located at the bottom of the gravitational wells of their respective dark matter halos. About $\sim$20\% of the quasars are outliers with offsets up to $\sim$750\,pc. This can be linked to disturbed morphologies in the \cii\ emission, most likely due to ongoing or recent mergers (such interactions can displace the interstellar medium more easily than the central black holes). We find no correlation between the central surface brightness of the FIR emission and the bolometric luminosity of the (positionally coincident) accreting central supermassive black hole.

The star formation rate densities in the host galaxies, derived from the FIR measurements, peak at the galaxy centers and typically reach values between 100 and 1000\,\msunyr\,kpc$^{-2}$. These values are below the Eddington limit for star formation, but are similar to those found in local ULIRGs. These star-formation rate densities drop toward larger radii by typically an order of magnitude, reaching values that exceed those found in nearby galaxy disks. Likewise, the \cii/FIR ratio is lowest in the centers of the quasar hosts, where luminosity ratios as low as a few times $10^{-4}$ are found, similar to studies of high-redshift starburst galaxies that show similar far-infrared surface brightness densities. Toward the outskirts of the disks, these values increase by a factor of a few, also reflecting the fact that the \cii\ emission is typically more extended than the FIR continuum.

The observations presented here have typical on-source integration times of 0.5--1.5\,hr with ALMA. The brightness of the targets imply that even higher resolution imaging of the quasar hosts, pushing to a resolution well below 1\,kpc in the frame of the quasar host galaxies, is within reach \citep{ven19}. Amongst other things, such studies would then allow one to probe the formation of massive galaxies in the early universe in detail and to investigate the interplay between emission from the accreting supermassive black hole and the interstellar medium down to scales of $<$100\,pc.

\newpage

\acknowledgments 
We thank the referee for carefully reading the manuscript and providing constructive comments and suggestions.
B.P.V., F.W., Ml.N., Ma.N., and A.B.D.\ acknowledge funding through the ERC Advanced Grant 740246 (Cosmic Gas). 
This paper makes use of the following ALMA data: \\
ADS/JAO.ALMA\#2012.1.00240.S, \\
ADS/JAO.ALMA\#2012.1.00882.S, \\ 
ADS/JAO.ALMA\#2013.1.00273.S, \\
ADS/JAO.ALMA\#2015.1.00399.S, \\
ADS/JAO.ALMA\#2015.1.00692.S, \\
ADS/JAO.ALMA\#2016.1.00544.S, \\
ADS/JAO.ALMA\#2016.A.000018.S, \\
ADS/JAO.ALMA\#2017.1.00396.S, \\
ADS/JAO.ALMA\#2017.1.01301.S, \\
ADS/JAO.ALMA\#2018.1.00908.S.
ALMA is a partnership of ESO
(representing its member states), NSF (USA), and NINS (Japan), together
with NRC (Canada) and NSC and ASIAA (Taiwan), in cooperation with the
Republic of Chile. The Joint ALMA Observatory is operated by ESO,
AUI/NRAO, and NAOJ.
This work has made use of data from the European Space Agency (ESA) mission
Gaia (\url{https://www.cosmos.esa.int/gaia}), processed by the Gaia
Data Processing and Analysis Consortium (DPAC,
\url{https://www.cosmos.esa.int/web/gaia/dpac/consortium}). Funding for the DPAC
has been provided by national institutions, in particular the institutions
participating in the Gaia Multilateral Agreement.

\vspace{5mm}
\facilities{ALMA}

\appendix

\section{Notes on individual objects}
\label{sec:notes}

We here summarize some notes on individual objects, sorted by R.A.

\subsection{P009--10}

Quasar P009--10 was part of the ALMA Cycle 3 \cii\ survey presented in \citet{dec18} and is the most luminous \cii\ source in our sample. Higher resolution imaging was obtained in Cycle~5. The source shows very extended \cii\ emission with an estimated spatial FWHM of 4.9$\times$3.4\,kpc$^2$ based on a 2D fit to the \cii\ map. This quasar has a disturbed \cii\ morphology (Figure~\ref{fig:allmaps}) and might undergo a merger with a very nearby source. 

\subsection{J0100+2802}

This luminous quasar is powered by the most massive black hole currently known at $z\gtrsim6$ of $>10^{10}$\,\msun\ \citep{wu15}. Observations of FIR, \cii, and CO emission from the host galaxy were presented in \citet{wan16}. The high-resolution data used in this paper were already presented in \citet{fwan19a}. 
Our line search revealed a potential companion galaxy $\sim$90\,kpc from the quasar (see Table~\ref{tab:companions} and the discussion in Section~\ref{sec:clustering}).

\subsection{J0109--3047}

The quasar J0109--3047 was imaged with ALMA in Cycle 1 \citep{ven16}. The quasar was targeted at kiloparsec resolution twice, in Cycle 2 and Cycle 3. We combined the data from both cycles to create deeper maps. Interestingly, for this source, the rest-frame UV emission lines from the broad-line region near the accreting black hole show very large blueshifts compared to the systemic redshift of the quasar as traced by \cii, with the blueshifts among the largest found in high-redshift quasars \citep{ven16,maz17b}. This could be due to an outflow or indicate that the accreting black hole is offset from the center of the host galaxy as traced by \cii\ emission, due to, for example, a recent major merger. This latter scenario is strengthened by the $\sim$700\,pc offset between the center of dust emission and the quasar (see Section~\ref{sec:location} and Figure~\ref{fig:position}).

\subsection{J0305--3150}

Low-resolution \cii\ observations of this quasar were published in \citet{ven16}. In ALMA Cycle 2 and Cycle 3, higher resolution observations were obtained. We combined the two high-resolution datasets in this paper. More recently, this quasar host was imaged at $\sim$400\,pc (0\farcs07) resolution \citep{ven19}. At 0\farcs07 resolution, the \cii\ morphology is highly irregular, and several faint companion galaxies are located within a few kiloparsec \citep{ven19}. The brightness of the source in Table\,\ref{tab:results} is taken from the higher S/N data in \citet{ven19}. We here only include the $\sim$1\,kpc resolution data for consistency. There are three companion \cii\ emitters in the field, all within 40\,kpc and 400\,\kms\ of the quasar \citep{ven19}. These line emitters were also found in our line search (Section~\ref{sec:lineemitters}). A fourth \cii\ emission line candidate we discovered in this field is at least $>$3000\,\kms\ away from the quasar and is likely not associated with it.

\subsection{P065--26}

The high-resolution FIR observations of quasar P065--26 show very extended \cii\ emission ($\sim$5\,kpc; see Table~\ref{tab:size}). The emission has a disturbed morphology (Figure~\ref{fig:allmaps}), an indication that 
the source is likely undergoing a merging event.

\subsection{J0842+1218}

In the field of this quasar, \citet{dec17} reported a companion \cii\ emitter at a distance of $\sim$47\,kpc. The $\sim$0\farcs25 resolution follow-up imaging of the quasar and the companion was recently published by \citet{nee19}. They also discovered a second, fainter companion about 31\,kpc from the quasar, which was rediscovered in our blind search for line emitters in this field (see Section~\ref{sec:lineemitters}).

\subsection{P167--13}

The \cii\ emission in this quasar at $z=6.5$ was independently discovered by \citet{wil17} and \citet{dec18}. The 0\farcs5 imaging by \citet{wil17} showed that this quasar has a nearby companion (5\,kpc from the quasar). This companion source was confirmed in higher-spatial-resolution ALMA data by \citet{nee19} and was also found in our blind line search (Section~\ref{sec:lineemitters}). The companion has also been detected in the rest-frame UV \citep{maz17b} and possibly in X-rays \citep{vit19}. 

\subsection*{J1120+0641}

The FIR emission of the host galaxy was initially detected by the Plateau the Bure Interferometer (PdBI) and published in \citet{ven12}. Follow-up higher spatial resolution imaging of the \cii\ line was obtained with ALMA in Cycle 1 \citep{ven17a}. In \citet{ven17a} a possible offset of 0\farcs5 between the accreting black hole and the host galaxy was reported. With the updated quasar position, derived by exploiting {\em Gaia} astrometry of nearby stars (Section~\ref{sec:astrometry}) and listed in Table~\ref{tab:sample}, the offset is now negligible (0\farcs03$\pm$0\farcs04). 

\subsection{P183+05}

The quasar P183+05 is the second brightest of our sample in both \cii\ and dust emission. Our line search revealed two candidate line emitters in the field (Table~\ref{tab:linecandidates}), one of which is potentially a \cii-emitting galaxy that is 140\,\kms\ and 57\,kpc from the quasar (Table~\ref{tab:companions}).

\subsection{J1306+0356}

This source was part of the low-resolution ALMA survey of $z\sim6$ quasar hosts published by \citet{dec18}. The low-resolution data showed that the quasar host is very extended, with a \cii\ size of $>$7\,kpc \citep{dec18}. High-resolution ALMA data revealed a more compact quasar host with a nearby ($\sim$5\,kpc) companion \citep{nee19}.

\subsection{J1319+0950}

This source was one of the first quasars imaged with ALMA in Cycle 0 \citep{wan13}. Higher-spatial-resolution 0\farcs3 observations were obtained in Cycle 1. The 0\farcs3 imaging of the \cii\ emission showed a velocity gradient consistent with a rotating gas disk \citep{sha17}. Reanalysis of these data reveals a very close companion only $\sim$3\,kpc from the quasar (see Figure~\ref{fig:allmaps}). Another possible companion is located 63\,kpc away (Table~\ref{tab:companions}).

\subsection{J1342+0928}

This is currently the highest-redshift quasar known. The first detection of the \cii\ and dust continuum emission from the host galaxy at $z=7.54$ was made with NOEMA \citep{ven17c}. The higher-spatial-resolution \cii\ imaging with ALMA described in this paper was recently published by \citet{ban19}, who classify this object as a merger. In the data cube, we discovered a candidate \cii\ emitter that is 27\,kpc from the quasar (Tables~\ref{tab:linecandidates} and \ref{tab:companions}).

\subsection{P231--20}

The quasar P231--20 at $z=6.6$ has a gas-rich companion located 9\,kpc from the quasar (Figure~\ref{fig:allmaps}), which was originally discovered by \citet{dec17}. The high-spatial-resolution observations presented here were already analyzed by \citet{nee19}. They reported the second, fainter \cii\ companion at a distance of 14\,kpc (see Table~\ref{tab:linecandidates}). A possible third companion is located 57\,kpc and $\sim$1600\,\kms\ from the quasar (Table~\ref{tab:companions}). A fourth candidate line emitter is not associated with the quasar, based on the large velocity difference ($>$15,000\,\kms).

\subsection{P308--21}

The \cii\ emission of quasar P308--21 shows very extended tails, suggesting that this source is undergoing a merger \citep{dec17}. The deep, high-resolution \cii\ imaging of this quasar revealed a main emission component centered on the quasar and two extended components stretching $\sim$1500\,\kms\ \citep{dec19}. The \cii\ kinematics and morphology can be explained by the tidal stripping of a satellite galaxy by the quasar host, confirming that the object is undergoing a merger \citep{dec19}.

\subsection{J2054--0005}

The quasar J2054--0005 was first observed by ALMA in Cycle 0. The high-spatial-resolution data presented here were obtained in Cycle 5. Our line search returned a potential companion galaxy at the same redshift as the quasar located at a projected distance of 77\,kpc.

\subsection{J2100--1715}

This is one of the sources that showed a FIR-bright companion at the same redshift in low-resolution data \citep{dec17}. The distance on the sky between the quasar and the companion is 61\,kpc \citep{dec17,nee19} and was rediscovered by our line search. 

\subsection{J2318--3113}

Both \cii\ and continuum emission of this quasar are very extended. The measured sizes are $\sim$5\,kpc, although the S/N of the detections is relatively low (peak S/N\,$\sim$\,5--6). There are two potential companion galaxies in this quasar field (Table~\ref{tab:companions}).

\subsection{J2348--3054}

Low-resolution \cii\ observations of this quasar at $z=6.9$ were obtained with ALMA in Cycle 1 and published by \citet{ven16}.  Higher-spatial-resolution imaging was obtained in Cycle 2 and Cycle 3. Because of possible astrometric errors in the Cycle 2 data, we did not combine the data from the two ALMA cycles, and we here only present the observations obtained in Cycle 3. Both \cii\ and dust emission are very compact, with deconvolved sizes $<$1\,kpc.

\section{Candidate \cii\ companions in the quasar fields}
\label{appendix}

We here present the new candidate \cii\ companions discussed in Section~\ref{sec:clustering} and summarized in Table~\ref{tab:companions}. These candidates were recovered in the blind line search and could potentially be \cii-emitting galaxies within 2000\,\kms\ of the quasar (Section~\ref{sec:clustering}). Single-pixel spectra, extracted at the peak of the candidate line emission, are shown toward the right in Figure~\ref{fig:linecandidatemaps}. Of the three maps that are shown, the central one displays the candidate emission line collapsed over 1.2\,$\times$\,FWHM. The maps toward the left and the right show emission away from the center, as indicated by the gray regions in the spectra toward the right. Contours in the maps start at $\pm$\,2\,$\sigma$ and increase in steps of $1\sigma$; positive contours are shown as black lines, and negative contours as red lines. The synthesized beam sizes are given in the bottom left corner of the maps. Gaussian fits are shown as blue lines on top of the spectra. The apertures used to extract the total spectrum of the candidates are indicated with light green circles. The parameters of the Gaussian fit to the total spectra are listed in Table~\ref{tab:linecandidates}. 
 
\begin{figure*}
\begin{center}
\includegraphics[width=\textwidth]{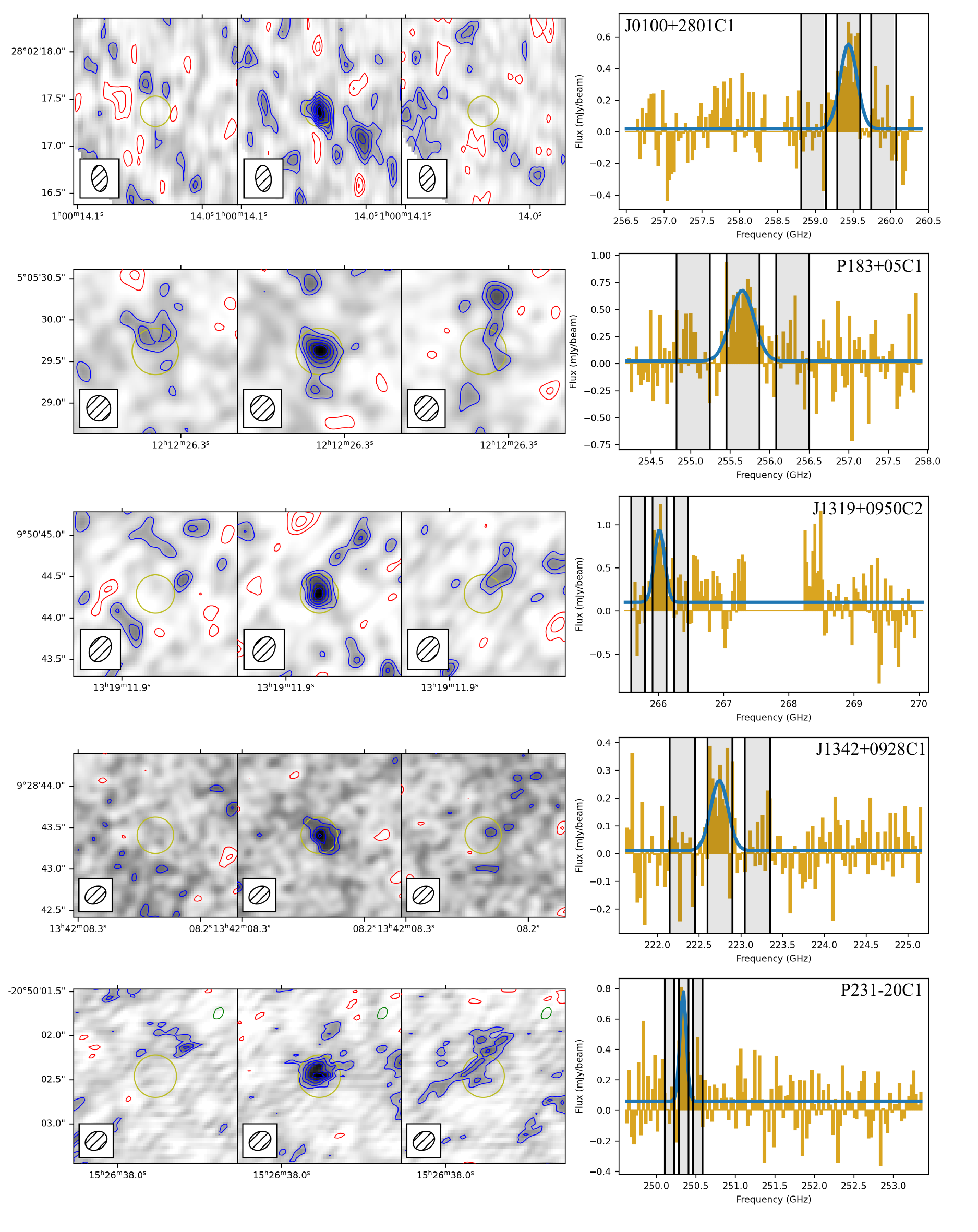}
\caption{Maps and spectra of the new candidate \cii\ companions found in the fields of the quasars, as discussed in Section~\ref{sec:lineemitters} and listed in Table~\ref{tab:companions}. See Appendix~\ref{appendix} for details.} 
\end{center}
\end{figure*}

\begin{figure*}\ContinuedFloat
\begin{center}
\includegraphics[width=\textwidth]{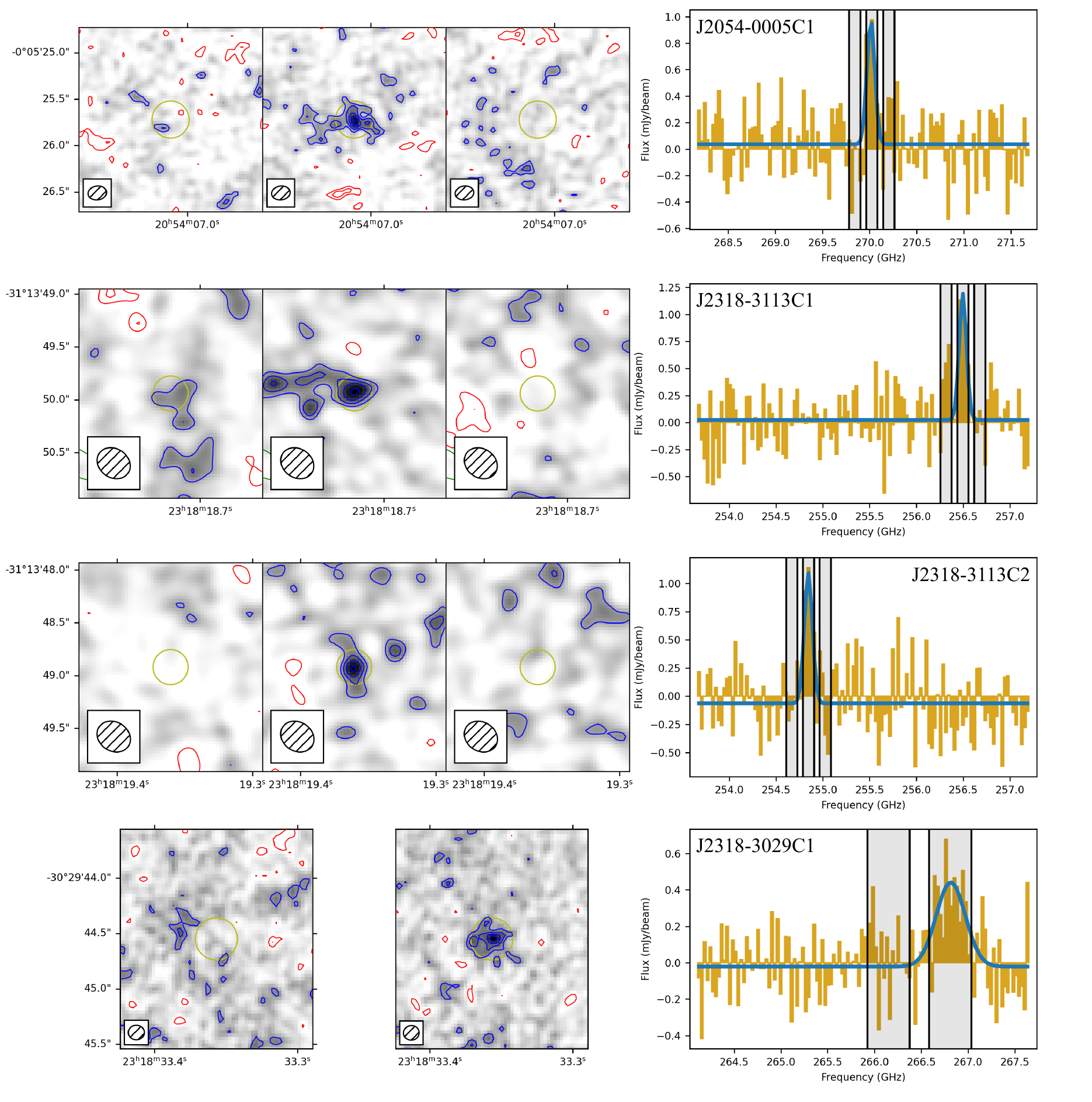}
\caption{Continued. \label{fig:linecandidatemaps}
}
\end{center}
\end{figure*}


\begin{thebibliography}{}
\expandafter\ifx\csname natexlab\endcsname\relax\def\natexlab#1{#1}\fi
\providecommand{\url}[1]{\href{#1}{#1}}

\bibitem[{{Ba{\~n}ados} {et~al.}(2018){Ba{\~n}ados}, {Venemans},
  {Mazzucchelli}, {Farina}, {Walter}, {Wang}, {Decarli}, {Stern}, {Fan},
  {Davies}, {Hennawi}, {Simcoe}, {Turner}, {Rix}, {Yang}, {Kelson}, {Rudie}, \&
  {Winters}}]{ban18a}
{Ba{\~n}ados}, E., {Venemans}, B.~P., {Mazzucchelli}, C., {et~al.} 2018,
  Nature, 553, 473

\bibitem[{{Ba{\~n}ados} {et~al.}(2019){Ba{\~n}ados}, {Novak}, {Neeleman},
  {Walter}, {Decarli}, {Venemans}, {Mazzucchelli}, {Carilli}, {Wang}, {Fan},
  {Farina}, \& {Rix}}]{ban19}
{Ba{\~n}ados}, E., {Novak}, M., {Neeleman}, M., {et~al.} 2019, ApJL, 881, L23

\bibitem[{{Bartlett} {et~al.}(2020){Bartlett}, {Desmond}, {Devriendt},
  {Ferreira}, \& {Slyz}}]{bar20}
{Bartlett}, D.~J., {Desmond}, H., {Devriendt}, J., {Ferreira}, P.~G., \&
  {Slyz}, A. 2020, arXiv e-prints, arXiv:2007.01353

\bibitem[{{Beelen} {et~al.}(2006){Beelen}, {Cox}, {Benford}, {Dowell},
  {Kov{\'a}cs}, {Bertoldi}, {Omont}, \& {Carilli}}]{bee06}
{Beelen}, A., {Cox}, P., {Benford}, D.~J., {et~al.} 2006, ApJ, 642, 694

\bibitem[{{Capak} {et~al.}(2015){Capak}, {Carilli}, {Jones}, {Casey},
  {Riechers}, {Sheth}, {Carollo}, {Ilbert}, {Karim}, {Lefevre}, {Lilly},
  {Scoville}, {Smolcic}, \& {Yan}}]{cap15}
{Capak}, P.~L., {Carilli}, C., {Jones}, G., {et~al.} 2015, \nat, 522, 455

\bibitem[{{Chambers} {et~al.}(2016){Chambers}, {Magnier}, {Metcalfe},
  {Flewelling}, {Huber}, {Waters}, {Denneau}, {Draper}, {Farrow}, {Finkbeiner},
  {Holmberg}, {Koppenhoefer}, {Price}, {Rest}, {Saglia}, {Schlafly}, {Smartt},
  {Sweeney}, {Wainscoat}, {Burgett}, {Chastel}, {Grav}, {Heasley}, {Hodapp},
  {Jedicke}, {Kaiser}, {Kudritzki}, {Luppino}, {Lupton}, {Monet}, {Morgan},
  {Onaka}, {Shiao}, {Stubbs}, {Tonry}, {White}, {Ba{\~n}ados}, {Bell},
  {Bender}, {Bernard}, {Boegner}, {Boffi}, {Botticella}, {Calamida},
  {Casertano}, {Chen}, {Chen}, {Cole}, {Deacon}, {Frenk}, {Fitzsimmons},
  {Gezari}, {Gibbs}, {Goessl}, {Goggia}, {Gourgue}, {Goldman}, {Grant},
  {Grebel}, {Hambly}, {Hasinger}, {Heavens}, {Heckman}, {Henderson}, {Henning},
  {Holman}, {Hopp}, {Ip}, {Isani}, {Jackson}, {Keyes}, {Koekemoer}, {Kotak},
  {Le}, {Liska}, {Long}, {Lucey}, {Liu}, {Martin}, {Masci}, {McLean}, {Mindel},
  {Misra}, {Morganson}, {Murphy}, {Obaika}, {Narayan}, {Nieto-Santisteban},
  {Norberg}, {Peacock}, {Pier}, {Postman}, {Primak}, {Rae}, {Rai}, {Riess},
  {Riffeser}, {Rix}, {R{\"o}ser}, {Russel}, {Rutz}, {Schilbach}, {Schultz},
  {Scolnic}, {Strolger}, {Szalay}, {Seitz}, {Small}, {Smith}, {Soderblom},
  {Taylor}, {Thomson}, {Taylor}, {Thakar}, {Thiel}, {Thilker}, {Unger},
  {Urata}, {Valenti}, {Wagner}, {Walder}, {Walter}, {Watters}, {Werner},
  {Wood-Vasey}, \& {Wyse}}]{cha16}
{Chambers}, K.~C., {Magnier}, E.~A., {Metcalfe}, N., {et~al.} 2016, arXiv
  e-prints, arXiv:1612.05560

\bibitem[Chanial et al.(2007)]{cha07} 
{Chanial}, P., {Flores}, H., {Guiderdoni}, B., {et~al.} 2007, A\&A, 462, 81

\bibitem[{{da Cunha} {et~al.}(2013){da Cunha}, {Groves}, {Walter}, {Decarli},
  {Weiss}, {Bertoldi}, {Carilli}, {Daddi}, {Elbaz}, {Ivison}, {Maiolino},
  {Riechers}, {Rix}, {Sargent}, \& {Smail}}]{dac13}
{da Cunha}, E., {Groves}, B., {Walter}, F., {et~al.} 2013, ApJ, 766, 13

\bibitem[{{De Rosa} {et~al.}(2014){De Rosa}, {Venemans}, {Decarli}, {Gennaro},
  {Simcoe}, {Dietrich}, {Peterson}, {Walter}, {Frank}, {McMahon}, {Hewett},
  {Mortlock}, \& {Simpson}}]{der14}
{De Rosa}, G., {Venemans}, B.~P., {Decarli}, R., {et~al.} 2014, ApJ, 790, 145

\bibitem[{{Decarli} {et~al.}(2017){Decarli}, {Walter}, {Venemans},
  {Ba{\~n}ados}, {Bertoldi}, {Carilli}, {Fan}, {Farina}, {Mazzucchelli},
  {Riechers}, {Rix}, {Strauss}, {Wang}, \& {Yang}}]{dec17}
{Decarli}, R., {Walter}, F., {Venemans}, B.~P., {et~al.} 2017, Nature, 545, 457

\bibitem[{{Decarli} {et~al.}(2018){Decarli}, {Walter}, {Venemans},
  {Ba{\~n}ados}, {Bertoldi}, {Carilli}, {Fan}, {Farina}, {Mazzucchelli},
  {Riechers}, {Rix}, {Strauss}, {Wang}, \& {Yang}}]{dec18}
---. 2018, ApJ, 854, 97

\bibitem[{{Decarli} {et~al.}(2019){Decarli}, {Dotti}, {Ba{\~n}ados}, {Farina},
  {Walter}, {Carilli}, {Fan}, {Mazzucchelli}, {Neeleman}, {Novak}, {Riechers},
  {Strauss}, {Venemans}, {Yang}, \& {Wang}}]{dec19}
{Decarli}, R., {Dotti}, M., {Ba{\~n}ados}, E., {et~al.} 2019, ApJ, 880, 157

\bibitem[Decarli et al.(2020)]{dec20} 
Decarli, R., Aravena, M., Boogaard, L., et al.\ 2020, arXiv:2009.10744

\bibitem[{{D{\'{\i}}az-Santos} {et~al.}(2013){D{\'{\i}}az-Santos}, {Armus},
  {Charmandaris}, {Stierwalt}, {Murphy}, {Haan}, {Inami}, {Malhotra},
  {Meijerink}, {Stacey}, {Petric}, {Evans}, {Veilleux}, {van der Werf}, {Lord},
  {Lu}, {Howell}, {Appleton}, {Mazzarella}, {Surace}, {Xu}, {Schulz},
  {Sanders}, {Bridge}, {Chan}, {Frayer}, {Iwasawa}, {Melbourne}, \&
  {Sturm}}]{dia13}
{D{\'{\i}}az-Santos}, T., {Armus}, L., {Charmandaris}, V., {et~al.} 2013, ApJ,
  774, 68
  
\bibitem[{{D{\'{\i}}az-Santos} {et~al.}(2014){D{\'{\i}}az-Santos}, {Armus},
  {Charmandaris}, {Stierwalt}, {Murphy}, {Haan}, {Inami}, {Malhotra},
  {Meijerink}, {Stacey}, {Petric}, {Evans}, {Veilleux}, {van der Werf}, {Lord},
  {Lu}, {Howell}, {Appleton}, {Mazzarella}, {Surace}, {Xu}, {Schulz},
  {Sanders}, {Bridge}, {Chan}, {Frayer}, {Iwasawa}, {Melbourne}, \&
  {Sturm}}]{dia14}
---. 2014, ApJL, 788, L17

\bibitem[{{D{\'\i}az-Santos} {et~al.}(2017){D{\'\i}az-Santos}, {Armus},
  {Charmandaris}, {Lu}, {Stierwalt}, {Stacey}, {Malhotra}, {van der Werf},
  {Howell}, {Privon}, {Mazzarella}, {Goldsmith}, {Murphy}, {Barcos-Mu{\~n}oz},
  {Linden}, {Inami}, {Larson}, {Evans}, {Appleton}, {Iwasawa}, {Lord},
  {Sanders}, \& {Surace}}]{dia17}
---. 2017, ApJ, 846, 32

\bibitem[{{Downes} \& {Solomon}(1998)}]{dow98}
{Downes}, D., \& {Solomon}, P.~M. 1998, ApJ, 507, 615

\bibitem[{{Edge} {et~al.}(2013){Edge}, {Sutherland}, {Kuijken}, {Driver},
  {McMahon}, {Eales}, \& {Emerson}}]{edg13}
{Edge}, A., {Sutherland}, W., {Kuijken}, K., {et~al.} 2013, The Messenger, 154,
  32

\bibitem[{{Ellison} {et~al.}(2010){Ellison}, {Prochaska}, {Hennawi}, {Lopez},
  {Usher}, {Wolfe}, {Russell}, \& {Benn}}]{ell10}
{Ellison}, S.~L., {Prochaska}, J.~X., {Hennawi}, J., {et~al.} 2010, \mnras,
  406, 1435

\bibitem[{{Gaia Collaboration} {et~al.}(2018){Gaia Collaboration}, {Brown},
  {Vallenari}, {Prusti}, {de Bruijne}, {Babusiaux}, {Bailer-Jones}, {Biermann},
  {Evans}, {Eyer}, \& et~al.}]{gaia18}
{Gaia Collaboration}, {Brown}, A.~G.~A., {Vallenari}, A., {et~al.} 2018, A\&A,
  616, A1

\bibitem[{{Gonz{\'a}lez-L{\'o}pez} {et~al.}(2019){Gonz{\'a}lez-L{\'o}pez},
  {Decarli}, {Pavesi}, {Walter}, {Aravena}, {Carilli}, {Boogaard}, {Popping},
  {Weiss}, {Assef}, {Bauer}, {Bertoldi}, {Bouwens}, {Contini}, {Cortes}, {Cox},
  {da Cunha}, {Daddi}, {D{\'\i}az-Santos}, {Inami}, {Hodge}, {Ivison}, {Le
  F{\`e}vre}, {Magnelli}, {Oesch}, {Riechers}, {Rix}, {Smail}, {Swinbank},
  {Somerville}, {Uzgil}, \& {van der Werf}}]{gon19}
{Gonz{\'a}lez-L{\'o}pez}, J., {Decarli}, R., {Pavesi}, R., {et~al.} 2019, ApJ,
  882, 139

\bibitem[{{Gullberg} {et~al.}(2018){Gullberg}, {Swinbank}, {Smail}, {Biggs},
  {Bertoldi}, {De Breuck}, {Chapman}, {Chen}, {Cooke}, {Coppin}, {Cox},
  {Dannerbauer}, {Dunlop}, {Edge}, {Farrah}, {Geach}, {Greve}, {Hodge}, {Ibar},
  {Ivison}, {Karim}, {Schinnerer}, {Scott}, {Simpson}, {Stach}, {Thomson}, {van
  der Werf}, {Walter}, {Wardlow}, \& {Weiss}}]{gul18}
{Gullberg}, B., {Swinbank}, A.~M., {Smail}, I., {et~al.} 2018, ApJ, 859, 12

\bibitem[{{Helou} {et~al.}(1988){Helou}, {Khan}, {Malek}, \& {Boehmer}}]{hel88}
{Helou}, G., {Khan}, I.~R., {Malek}, L., \& {Boehmer}, L. 1988, ApJS, 68, 151

\bibitem[{{Hennawi} {et~al.}(2006){Hennawi}, {Strauss}, {Oguri}, {Inada},
  {Richards}, {Pindor}, {Schneider}, {Becker}, {Gregg}, {Hall}, {Johnston},
  {Fan}, {Burles}, {Schlegel}, {Gunn}, {Lupton}, {Bahcall}, {Brunner}, \&
  {Brinkmann}}]{hen06}
{Hennawi}, J.~F., {Strauss}, M.~A., {Oguri}, M., {et~al.} 2006, AJ, 131, 1

\bibitem[{{Herrera-Camus} {et~al.}(2018){Herrera-Camus}, {Sturm},
  {Graci{\'a}-Carpio}, {Lutz}, {Contursi}, {Veilleux}, {Fischer},
  {Gonz{\'a}lez-Alfonso}, {Poglitsch}, {Tacconi}, {Genzel}, {Maiolino},
  {Sternberg}, {Davies}, \& {Verma}}]{her18}
{Herrera-Camus}, R., {Sturm}, E., {Graci{\'a}-Carpio}, J., {et~al.} 2018, ApJ,
  861, 95

\bibitem[{{Hodge} {et~al.}(2016){Hodge}, {Swinbank}, {Simpson}, {Smail},
  {Walter}, {Alexander}, {Bertoldi}, {Biggs}, {Brandt}, {Chapman}, {Chen},
  {Coppin}, {Cox}, {Dannerbauer}, {Edge}, {Greve}, {Ivison}, {Karim},
  {Knudsen}, {Menten}, {Rix}, {Schinnerer}, {Wardlow}, {Weiss}, \& {van der
  Werf}}]{hod16}
{Hodge}, J.~A., {Swinbank}, A.~M., {Simpson}, J.~M., {et~al.} 2016, ApJ, 833,
  103

\bibitem[{{Kaczmarczik} {et~al.}(2009){Kaczmarczik}, {Richards}, {Mehta}, \&
  {Schlegel}}]{kac09}
{Kaczmarczik}, M.~C., {Richards}, G.~T., {Mehta}, S.~S., \& {Schlegel}, D.~J.
  2009, AJ, 138, 19

\bibitem[{{Kennicutt} \& {Evans}(2012)}]{ken12}
{Kennicutt}, R.~C., \& {Evans}, N.~J. 2012, ARA\&A, 50, 531

\bibitem[{{Kroupa} \& {Weidner}(2003)}]{kro03}
{Kroupa}, P., \& {Weidner}, C. 2003, ApJ, 598, 1076

\bibitem[{{Lamarche} {et~al.}(2018){Lamarche}, {Verma}, {Vishwas}, {Stacey},
  {Brisbin}, {Ferkinhoff}, {Nikola}, {Higdon}, {Higdon}, \& {Tecza}}]{lam18}
{Lamarche}, C., {Verma}, A., {Vishwas}, A., {et~al.} 2018, ApJ, 867, 140

\bibitem[{{Leipski} {et~al.}(2014){Leipski}, {Meisenheimer}, {Walter}, {Klaas},
  {Dannerbauer}, {De Rosa}, {Fan}, {Haas}, {Krause}, \& {Rix}}]{lei14}
{Leipski}, C., {Meisenheimer}, K., {Walter}, F., {et~al.} 2014, ApJ, 785, 154

\bibitem[{{Leroy} {et~al.}(2013){Leroy}, {Walter}, {Sandstrom}, {Schruba},
  {Munoz-Mateos}, {Bigiel}, {Bolatto}, {Brinks}, {de Blok}, {Meidt}, {Rix},
  {Rosolowsky}, {Schinnerer}, {Schuster}, \& {Usero}}]{ler13}
{Leroy}, A.~K., {Walter}, F., {Sandstrom}, K., {et~al.} 2013, AJ, 146, 19

\bibitem[{{Litke} {et~al.}(2019){Litke}, {Marrone}, {Spilker}, {Aravena},
  {B{\'e}thermin}, {Chapman}, {Chen}, {de Breuck}, {Dong}, {Gonzalez}, {Greve},
  {Hayward}, {Hezaveh}, {Jarugula}, {Ma}, {Morningstar}, {Narayanan}, {Phadke},
  {Reuter}, {Vieira}, \& {Weiss}}]{lit19}
{Litke}, K.~C., {Marrone}, D.~P., {Spilker}, J.~S., {et~al.} 2019, ApJ, 870, 80

\bibitem[{{Lutz} {et~al.}(2016){Lutz}, {Berta}, {Contursi}, {F{\"o}rster
  Schreiber}, {Genzel}, {Graci{\'a}-Carpio}, {Herrera-Camus}, {Netzer},
  {Sturm}, {Tacconi}, {Tadaki}, \& {Veilleux}}]{lut16}
{Lutz}, D., {Berta}, S., {Contursi}, A., {et~al.} 2016, A\&A, 591, A136

\bibitem[{{Marian} {et~al.}(2019){Marian}, {Jahnke}, {Mechtley}, {Cohen},
  {Husemann}, {Jones}, {Koekemoer}, {Schulze}, {van der Wel}, {Villforth}, \&
  {Windhorst}}]{mar19}
{Marian}, V., {Jahnke}, K., {Mechtley}, M., {et~al.} 2019, ApJ, 882, 141

\bibitem[{{Mazzucchelli} {et~al.}(2017){Mazzucchelli}, {Ba{\~n}ados},
  {Venemans}, {Decarli}, {Farina}, {Walter}, {Eilers}, {Rix}, {Simcoe},
  {Stern}, {Fan}, {Schlafly}, {De Rosa}, {Hennawi}, {Chambers}, {Greiner},
  {Burgett}, {Draper}, {Kaiser}, {Kudritzki}, {Magnier}, {Metcalfe}, {Waters},
  \& {Wainscoat}}]{maz17b}
{Mazzucchelli}, C., {Ba{\~n}ados}, E., {Venemans}, B.~P., {et~al.} 2017, ApJ,
  849, 91

\bibitem[{{McMullin} {et~al.}(2007){McMullin}, {Waters}, {Schiebel}, {Young},
  \& {Golap}}]{mul07}
{McMullin}, J.~P., {Waters}, B., {Schiebel}, D., {Young}, W., \& {Golap}, K.
  2007, in Astronomical Society of the Pacific Conference Series, Vol. 376,
  Astronomical Data Analysis Software and Systems XVI, ed. R.~A. {Shaw},
  F.~{Hill}, \& D.~J. {Bell}, 127

\bibitem[{{Murphy} {et~al.}(2011){Murphy}, {Condon}, {Schinnerer}, {Kennicutt},
  {Calzetti}, {Armus}, {Helou}, {Turner}, {Aniano}, {Beir{\~a}o}, {Bolatto},
  {Brandl}, {Croxall}, {Dale}, {Donovan Meyer}, {Draine}, {Engelbracht},
  {Hunt}, {Hao}, {Koda}, {Roussel}, {Skibba}, \& {Smith}}]{mur11}
{Murphy}, E., {Condon}, J., {Schinnerer}, E., {et~al.} 2011, ApJ, 737, 67

\bibitem[{{Neeleman} {et~al.}(2019){Neeleman}, {Ba{\~n}ados}, {Walter},
  {Decarli}, {Venemans}, {Carilli}, {Fan}, {Farina}, {Mazzucchelli}, {Novak},
  {Riechers}, {Rix}, \& {Wang}}]{nee19}
{Neeleman}, M., {Ba{\~n}ados}, E., {Walter}, F., {et~al.} 2019, ApJ, 882, 10

\bibitem[{{Neeleman} {et~al.}(2020){Neeleman}, {Walter}, {Novak}, {Venemans},
  {Decarli}, {Carilli}, {Fan}, {Farina}, {Mazzucchelli}, {Riechers}, {Rix}, \&
  {Wang}}]{nee20}
{Neeleman}, M., {Walter}, F., {Novak}, M., {et~al.} 2020, ApJ, in prep

\bibitem[{{Nguyen} {et~al.}(2020){Nguyen}, {Lira}, {Trakhtenbrot}, {Netzer},
  {Cicone}, {Maiolino}, \& {Shemmer}}]{ngu20}
{Nguyen}, N.~H., {Lira}, P., {Trakhtenbrot}, B., {et~al.} 2020, ApJ, 895, 74

\bibitem[{{Novak} {et~al.}(2019){Novak}, {Ba{\~n}ados}, {Decarli}, {Walter},
  {Venemans}, {Neeleman}, {Farina}, {Mazzucchelli}, {Carilli}, {Fan}, {Rix}, \&
  {Wang}}]{nov19}
{Novak}, M., {Ba{\~n}ados}, E., {Decarli}, R., {et~al.} 2019, ApJ, 881, 63

\bibitem[{{Novak} {et~al.}(2020){Novak}, {Venemans}, {Walter}, {Neeleman},
  {Ba{\~n}ados}, {Decarli}, {Farina}, {Mazzucchelli}, {Carilli}, {Fan}, {Rix},
  \& {Wang}}]{nov20}
{Novak}, M., {Venemans}, B., {Walter}, F., {et~al.} 2020, ApJ, submitted

\bibitem[{{Pensabene} {et~al.}(2020){Pensabene}, {Carniani}, {Perna}, {Cresci},
  {Decarli}, {Maiolino}, \& {Marconi}}]{pen20}
{Pensabene}, A., {Carniani}, S., {Perna}, M., {et~al.} 2020, A\&A, 637, A84

\bibitem[{{Planck Collaboration} {et~al.}(2016){Planck Collaboration}, {Ade},
  {Aghanim}, {Arnaud}, {Ashdown}, {Aumont}, {Baccigalupi}, {Banday},
  {Barreiro}, {Bartlett}, \& et~al.}]{pla16}
{Planck Collaboration}, {Ade}, P.~A.~R., {Aghanim}, N., {et~al.} 2016, A\&A,
  594, A13

\bibitem[{{Rybak} {et~al.}(2020){Rybak}, {Hodge}, {Vegetti}, {van der Werf},
  {Andreani}, {Graziani}, \& {McKean}}]{ryb20}
{Rybak}, M., {Hodge}, J.~A., {Vegetti}, S., {et~al.} 2020, MNRAS, 494, 5542

\bibitem[{{Rybak} {et~al.}(2019){Rybak}, {Calistro Rivera}, {Hodge}, {Smail},
  {Walter}, {van der Werf}, {da Cunha}, {Chen}, {Dannerbauer}, {Ivison},
  {Karim}, {Simpson}, {Swinbank}, \& {Wardlow}}]{ryb19}
{Rybak}, M., {Calistro Rivera}, G., {Hodge}, J.~A., {et~al.} 2019, ApJ, 876,
  112

\bibitem[{{Sargsyan} {et~al.}(2014){Sargsyan}, {Samsonyan}, {Lebouteiller},
  {Weedman}, {Barry}, {Bernard-Salas}, {Houck}, \& {Spoon}}]{sar14}
{Sargsyan}, L., {Samsonyan}, A., {Lebouteiller}, V., {et~al.} 2014, ApJ, 790,
  15

\bibitem[{{Schneider} {et~al.}(2015){Schneider}, {Bianchi}, {Valiante},
  {Risaliti}, \& {Salvadori}}]{sch15}
{Schneider}, R., {Bianchi}, S., {Valiante}, R., {Risaliti}, G., \& {Salvadori},
  S. 2015, A\&A, 579, A60

\bibitem[{{Shao} {et~al.}(2017){Shao}, {Wang}, {Jones}, {Carilli}, {Walter},
  {Fan}, {Riechers}, {Bertoldi}, {Wagg}, {Strauss}, {Omont}, {Cox}, {Jiang},
  {Narayanan}, \& {Menten}}]{sha17}
{Shao}, Y., {Wang}, R., {Jones}, G.~C., {et~al.} 2017, ApJ, 845, 138

\bibitem[{{Shen} {et~al.}(2019){Shen}, {Wu}, {Jiang}, {Ba{\~n}ados}, {Fan},
  {Ho}, {Riechers}, {Strauss}, {Venemans}, {Vestergaard}, {Walter}, {Wang},
  {Willott}, {Wu}, \& {Yang}}]{she19}
{Shen}, Y., {Wu}, J., {Jiang}, L., {et~al.} 2019, ApJ, 873, 35

\bibitem[{{Smith} {et~al.}(2017){Smith}, {Croxall}, {Draine}, {De Looze},
  {Sandstrom}, {Armus}, {Beir{\~a}o}, {Bolatto}, {Boquien}, {Brandl},
  {Crocker}, {Dale}, {Galametz}, {Groves}, {Helou}, {Herrera-Camus}, {Hunt},
  {Kennicutt}, {Walter}, \& {Wolfire}}]{smi17}
{Smith}, J.~D.~T., {Croxall}, K., {Draine}, B., {et~al.} 2017, ApJ, 834, 5

\bibitem[{{Trakhtenbrot} {et~al.}(2017){Trakhtenbrot}, {Lira}, {Netzer},
  {Cicone}, {Maiolino}, \& {Shemmer}}]{tra17}
{Trakhtenbrot}, B., {Lira}, P., {Netzer}, H., {et~al.} 2017, ApJ, 836, 8

\bibitem[{{Venemans}(2006)}]{ven06}
{Venemans}, B.~P. 2006, AN, 327, 196

\bibitem[{{Venemans} {et~al.}(2019){Venemans}, {Neeleman}, {Walter}, {Novak},
  {Decarli}, {Hennawi}, \& {Rix}}]{ven19}
{Venemans}, B.~P., {Neeleman}, M., {Walter}, F., {et~al.} 2019, \apj, 874, L30

\bibitem[{{Venemans} {et~al.}(2016){Venemans}, {Walter}, {Zschaechner},
  {Decarli}, {De Rosa}, {Findlay}, {McMahon}, \& {Sutherland}}]{ven16}
{Venemans}, B.~P., {Walter}, F., {Zschaechner}, L., {et~al.} 2016, ApJ, 816, 37

\bibitem[{{Venemans} {et~al.}(2007){Venemans}, {R{\"o}ttgering}, {Miley}, {van
  Breugel}, {de Breuck}, {Kurk}, {Pentericci}, {Stanford}, {Overzier}, {Croft},
  \& {Ford}}]{ven07a}
{Venemans}, B.~P., {R{\"o}ttgering}, H.~J.~A., {Miley}, G.~K., {et~al.} 2007,
  A\&A, 461, 823

\bibitem[{{Venemans} {et~al.}(2012){Venemans}, {McMahon}, {Walter}, {Decarli},
  {Cox}, {Neri}, {Hewett}, {Mortlock}, {Simpson}, \& {Warren}}]{ven12}
{Venemans}, B.~P., {McMahon}, R.~G., {Walter}, F., {et~al.} 2012, ApJL, 751,
  L25

\bibitem[{{Venemans} {et~al.}(2017{\natexlab{a}}){Venemans}, {Walter},
  {Decarli}, {Ba{\~n}ados}, {Hodge}, {Hewett}, {McMahon}, {Mortlock}, \&
  {Simpson}}]{ven17a}
{Venemans}, B.~P., {Walter}, F., {Decarli}, R., {et~al.} 2017{\natexlab{a}},
  ApJ, 837, 146

\bibitem[{{Venemans} {et~al.}(2017{\natexlab{b}}){Venemans}, {Walter},
  {Decarli}, {Ba{\~n}ados}, {Carilli}, {Winters}, {Schuster}, {da Cunha},
  {Fan}, {Farina}, {Mazzucchelli}, {Rix}, \& {Weiss}}]{ven17c}
---. 2017{\natexlab{b}}, ApJL, 851, L8

\bibitem[{{Venemans} {et~al.}(2018){Venemans}, {Decarli}, {Walter},
  {Ba{\~n}ados}, {Bertoldi}, {Fan}, {Farina}, {Mazzucchelli}, {Riechers},
  {Rix}, {Wang}, \& {Yang}}]{ven18}
{Venemans}, B.~P., {Decarli}, R., {Walter}, F., {et~al.} 2018, ApJ, 866, 159

\bibitem[{{Vito} {et~al.}(2019){Vito}, {Brandt}, {Bauer}, {Gilli}, {Luo},
  {Zamorani}, {Calura}, {Comastri}, {Mazzucchelli}, {Mignoli}, {Nanni},
  {Shemmer}, {Vignali}, {Brusa}, {Cappelluti}, {Civano}, \&
  {Volonteri}}]{vit19}
{Vito}, F., {Brandt}, W.~N., {Bauer}, F.~E., {et~al.} 2019, A\&A, 628, L6

\bibitem[{{Walter} {et~al.}(2009){Walter}, {Riechers}, {Cox}, {Neri},
  {Carilli}, {Bertoldi}, {Weiss}, \& {Maiolino}}]{wal09b}
{Walter}, F., {Riechers}, D., {Cox}, P., {et~al.} 2009, Nature, 457, 699

\bibitem[{{Wang} {et~al.}(2019){Wang}, {Wang}, {Fan}, {Wu}, {Yang}, {Neri}, \&
  {Yue}}]{fwan19a}
{Wang}, F., {Wang}, R., {Fan}, X., {et~al.} 2019, ApJ, 880, 2

\bibitem[{{Wang} {et~al.}(2011){Wang}, {Wagg}, {Carilli}, {Walter}, {Riechers},
  {Willott}, {Bertoldi}, {Omont}, {Beelen}, {Cox}, {Strauss}, {Bergeron},
  {Forveille}, {Menten}, \& {Fan}}]{wan11a}
{Wang}, R., {Wagg}, J., {Carilli}, C.~L., {et~al.} 2011, ApJL, 739, L34

\bibitem[{{Wang} {et~al.}(2013){Wang}, {Wagg}, {Carilli}, {Walter}, {Lentati},
  {Fan}, {Riechers}, {Bertoldi}, {Narayanan}, {Strauss}, {Cox}, {Omont},
  {Menten}, {Knudsen}, {Neri}, \& {Jiang}}]{wan13}
---. 2013, ApJ, 773, 44

\bibitem[{{Wang} {et~al.}(2016){Wang}, {Wu}, {Neri}, {Fan}, {Walter},
  {Carilli}, {Momjian}, {Bertoldi}, {Strauss}, {Li}, {Wang}, {Riechers},
  {Jiang}, {Omont}, {Wagg}, \& {Cox}}]{wan16}
{Wang}, R., {Wu}, X.-B., {Neri}, R., {et~al.} 2016, ApJ, 830, 53

\bibitem[{{Weinberger} {et~al.}(2017){Weinberger}, {Springel}, {Hernquist},
  {Pillepich}, {Marinacci}, {Pakmor}, {Nelson}, {Genel}, {Vogelsberger},
  {Naiman}, \& {Torrey}}]{wei17}
{Weinberger}, R., {Springel}, V., {Hernquist}, L., {et~al.} 2017, MNRAS, 465,
  3291

\bibitem[{{Willott} {et~al.}(2017){Willott}, {Bergeron}, \& {Omont}}]{wil17}
{Willott}, C.~J., {Bergeron}, J., \& {Omont}, A. 2017, ApJ, 850, 108

\bibitem[{{Willott} {et~al.}(2013){Willott}, {Omont}, \& {Bergeron}}]{wil13}
{Willott}, C.~J., {Omont}, A., \& {Bergeron}, J. 2013, ApJ, 770, 13

\bibitem[{{Wu} {et~al.}(2015){Wu}, {Wang}, {Fan}, {Yi}, {Zuo}, {Bian}, {Jiang},
  {McGreer}, {Wang}, {Yang}, {Yang}, {Thompson}, \& {Beletsky}}]{wu15}
{Wu}, X.-B., {Wang}, F., {Fan}, X., {et~al.} 2015, \nat, 518, 512

\bibitem[{{Yang} {et~al.}(2019){Yang}, {Venemans}, {Wang}, {Fan}, {Novak},
  {Decarli}, {Walter}, {Yue}, {Momjian}, {Keeton}, {Wang}, {Zabludoff}, {Wu},
  \& {Bian}}]{yan19}
{Yang}, J., {Venemans}, B., {Wang}, F., {et~al.} 2019, ApJ, 880, 153

\bibitem[{{Yang} {et~al.}(2020){Yang}, {Wang}, {Fan}, {Hennawi}, {Davies},
  {Yue}, {Banados}, {Wu}, {Venemans}, {Barth}, {Bian}, {Boutsia}, {Decarli},
  {Farina}, {Green}, {Jiang}, {Li}, {Mazzucchelli}, \& {Walter}}]{yan20}
{Yang}, J., {Wang}, F., {Fan}, X., {et~al.} 2020, ApJL, 897, L14

\end{thebibliography}
\end{document}